\documentclass[english,superscriptaddress,floatfix]{revtex4}
\usepackage[T1]{fontenc}
\usepackage[latin9]{inputenc}
\usepackage[pdftex]{color}
\usepackage{babel}
\usepackage{amsmath}
\usepackage{amssymb}
\usepackage[pdftex]{graphicx}
\usepackage{esint}
\usepackage[unicode=true,
 bookmarks=false,
 breaklinks=true,pdfborder={0 0 1},backref=section,colorlinks=true]
 {hyperref}
\hypersetup{
 pdfcreator={},pdfproducer={LaTeX with hyperref},linkcolor=blue,anchorcolor=blue,citecolor=blue,filecolor=red,menucolor=red,pagecolor=red,urlcolor=blue,pdfstartview=FitV,pdfhighlight=/I,pdfpagelayout=OneColumn,hypertexnames=true}
\usepackage{breakurl}

\makeatletter
\@ifundefined{textcolor}{}
{%
 \definecolor{BLACK}{gray}{0}
 \definecolor{WHITE}{gray}{1}
 \definecolor{RED}{rgb}{1,0,0}
 \definecolor{GREEN}{rgb}{0,1,0}
 \definecolor{BLUE}{rgb}{0,0,1}
 \definecolor{CYAN}{cmyk}{1,0,0,0}
 \definecolor{MAGENTA}{cmyk}{0,1,0,0}
 \definecolor{YELLOW}{cmyk}{0,0,1,0}
}

\usepackage{babel}

 \@ifundefined{textcolor}{}{%
  \definecolor{BLACK}{gray}{0}
  \definecolor{WHITE}{gray}{1}
  \definecolor{RED}{rgb}{1,0,0}
  \definecolor{GREEN}{rgb}{0,1,0}
  \definecolor{BLUE}{rgb}{0,0,1}
  \definecolor{CYAN}{cmyk}{1,0,0,0}
  \definecolor{MAGENTA}{cmyk}{0,1,0,0}
  \definecolor{YELLOW}{cmyk}{0,0,1,0}
 }

 \@ifundefined{textcolor}{}{
  \definecolor{BLACK}{gray}{0}
  \definecolor{WHITE}{gray}{1}
  \definecolor{RED}{rgb}{1,0,0}
  \definecolor{GREEN}{rgb}{0,1,0}
  \definecolor{BLUE}{rgb}{0,0,1}
  \definecolor{CYAN}{cmyk}{1,0,0,0}
  \definecolor{MAGENTA}{cmyk}{0,1,0,0}
  \definecolor{YELLOW}{cmyk}{0,0,1,0}
 }
 \@ifundefined{definecolor}{
 }{}
 \@ifundefined{definecolor}{color}{}

\usepackage{color}

\newcommand{\be}{\begin{equation}}
\newcommand{\ee}{\end{equation}}
\newcommand{\bea}{\begin{eqnarray}}
\newcommand{\eea}{\end{eqnarray}}
\newcommand{\bse}{\begin{subequations}}
\newcommand{\ese}{\end{subequations}}

\usepackage{color}% define colors
\definecolor{d_red}{cmyk}{0.00, 0.81, 1.00, 0.27}
\definecolor{d_orange}{cmyk}{0.00, 0.33, 1.00, 0.00}
\definecolor{d_blue}{cmyk}{0.78, 0.47, 0.00, 0.20}
\definecolor{d_lgreen}{cmyk}{0.07, 0.00, 0.79, 0.29}
\definecolor{d_green}{cmyk}{0.66, 0.00, 0.71, 0.56}
\definecolor{d_blue}{cmyk}{0.78, 0.47, 0.00, 0.20}
\definecolor{d_dblue}{cmyk}{0.91, 0.79, 0.00, 0.22}
\definecolor{d_pink}{cmyk}{0.0, 0.79, 0.37, 0.29}
\definecolor{d_purple}{cmyk}{0.16, 0.54, 0.00, 0.70}
\definecolor{d_paleblue}{cmyk}{0.669, 0.338, 0.00, 0.373}
\definecolor{d_dpaleblue}{cmyk}{0.441, 0.290, 0.00, 0.580}
\definecolor{d_brown}{cmyk}{0.0, 0.490, 0.930, 0.350}
\definecolor{d_turquoise}{cmyk}{0.630, 0.04, 0.0, 0.440}
\definecolor{KIT-green}{RGB}{0, 150,130}
\definecolor{KIT-blue}{RGB}{70,100,170}

\def\bmx{\begin{pmatrix}}
\def\emx{\end{pmatrix}}

\usepackage[figure,table]{hypcap}% to correct a problem with hyperref

\makeatother

\begin{document}

\title{Cooper pairing of incoherent electrons: an electron-phonon version
of the Sachdev-Ye-Kitaev model}

\author{Ilya Esterlis}

\affiliation{Department of Physics, Stanford University, Stanford, California
94305, USA}

\author{J{\"o}rg Schmalian}

\affiliation{Institute for Theory of Condensed Matter, Karlsruhe Institute of
Technology, Karlsruhe, Germany}

\affiliation{Institute for Solid State Physics, Karlsruhe Institute of Technology,
Karlsruhe, Germany}
\begin{abstract}
We introduce and solve a model of interacting electrons and phonons that
is a natural generalization of the Sachdev-Ye-Kitaev-model and that becomes
superconducting at low temperatures.
In the normal state two Non-Fermi liquid fixed points with distinct universal exponents emerge. 
At weak coupling superconductivity prevents the onset of low-temperature quantum criticality, reminiscent
of the behavior in several heavy-electron and iron-based materials.
At strong coupling, pairing of highly incoherent fermions sets in deep in the Non-Fermi liquid regime, a behavior qualitatively similar to that in underdoped cuprate superconductors.  The pairing of incoherent time-reversal partners is protected
by a mechanism similar to Anderson's theorem for disordered superconductors.
The superconducting ground state is characterized by coherent quasiparticle excitations and
higher-order bound states thereof, revealing that it
 is no longer an ideal gas of Cooper pairs, but a strongly coupled
pair fluid. The normal-state incoherency primarily acts to suppress the weight of the superconducting coherence peak and reduce the condensation energy. Based on this we expect strong superconducting fluctuations, in particular at strong coupling.
\end{abstract}
\maketitle

\section{Introduction}

Superconductivity is the ultimate fate of a Fermi liquid at low temperatures\cite{Cooper1956,Bardeen1957lett,Bardeen1957,Kohn1965}.
A key assumption that gives rise to this Cooper instability is that
the excitations of a Fermi liquid are slowly-decaying Landau quasiparticles
with the same quantum numbers as free fermions. The resulting superconducting
ground state can be understood as an ideal gas of Cooper pairs. Since
superconductivity occurs in many systems where such sharp excitations
are absent, the conditions for pairing of incoherent electrons is
an important open problem. The emergence of a
sharp superconducting coherence peak of small weight from a broad
and structureless normal-state spectrum is in fact one of the hallmarks
of the cuprate superconductors\cite{Dessau1991,ZXSchrieffer1997,Campuzano1996,Fedorov1999,Feng2000},
where the weight of the coherence peak was shown to be
strongly correlated with the superfluid stiffness and the condensation
energy\cite{Feng2000}.
Key questions in this context are: Can one form Cooper
pairs from completely incoherent fermions? Are there sharp quasi-particles
in such a superconductor? Is the Cooper pair fluid that emerges still
an ideal gas of pairs?  

\begin{figure}
\includegraphics[scale=0.5]{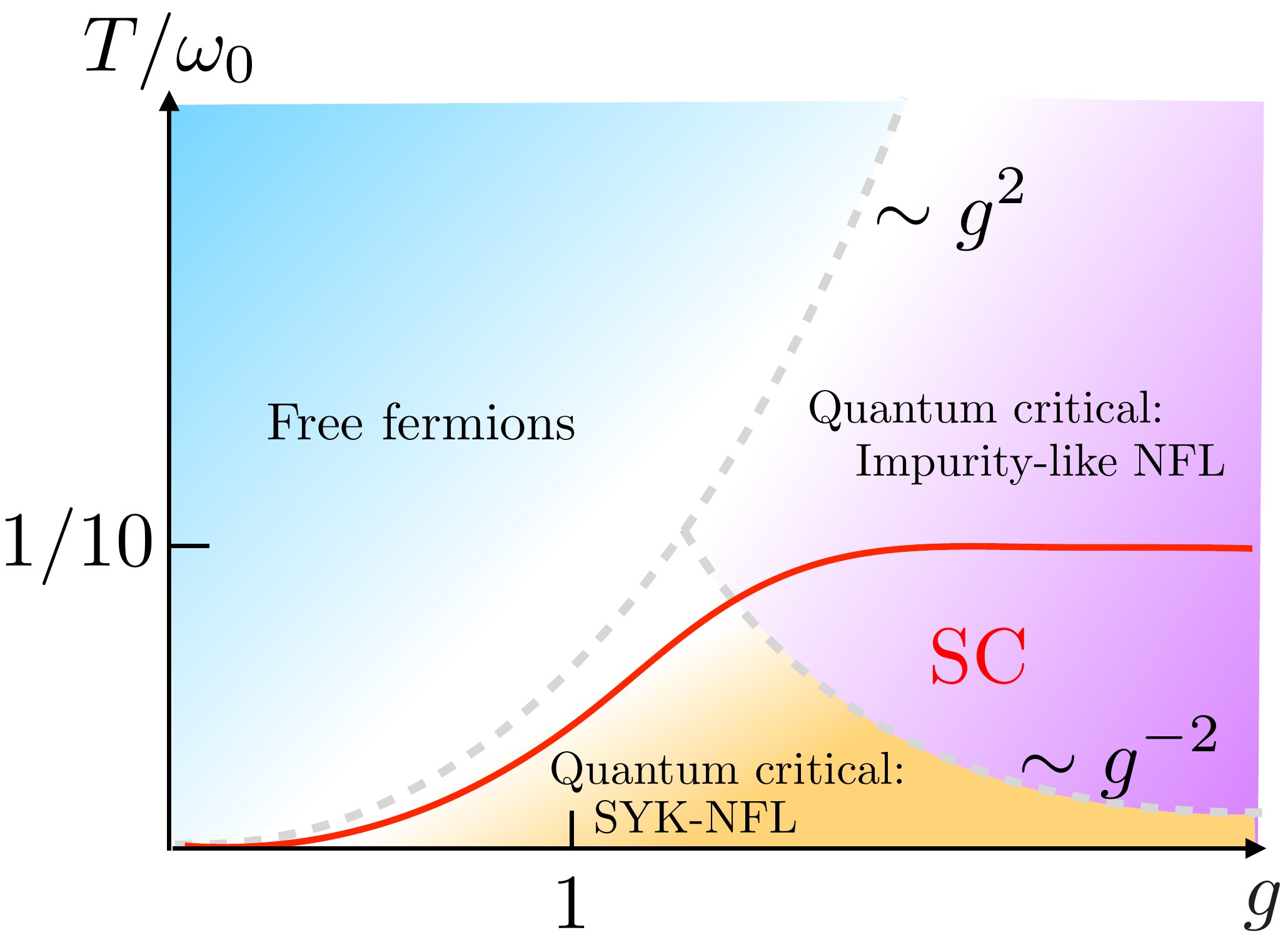}

\caption{Schematic phase diagram of the SYK-model for electron-boson coupling
as function of the dimensionless coupling constant $g=\bar{g}/\omega_{0}^{3/2}$,
where $\omega_{0}$ is the bare phonon frequency. At lowest $T$ the
normal state would be a Non-Fermi liquid state with anomalous exponents,
similar to other SYK models. For $g<1$ superconductivity sets in
at $T_{c}/\omega_{0}\propto g^{2}$, comparable to the temperature
where quantum critical SYK-NFL sets in. Thus pairing occurs instead
of the low-$T$ quantum critical state. At strong coupling a new intermediate-temperature
regime opens up that is characterized by fully incoherent fermions.
Coherent pairing of such incoherent fermions is still possible with
finite transition temperature $T_{c}\rightarrow0.112\omega_{0}$. }

\label{Fig: phase diagram schem}
\end{figure}

To address these questions in a theoretically well-controlled way
it is highly desirable to identify a solvable model for non-quasiparticle
superconductivity. A crucial issue is the proper interplay of Non-Fermi
liquid excitations and the pairing interaction. For example, the spectral
function of a Fermi liquid right at the Fermi surface,
\begin{equation}
A_{{\rm FL}}\left(\omega\right)=Z_{{\rm FL}}\delta\left(\omega\right),
\end{equation}
 is expected to transform for a quantum-critical system to the power-law
form 
\begin{equation}
A_{{\rm QC}}\left(\omega\right)=A_{0}\left|\omega\right|^{2\Delta-1}
\end{equation}
with exponent $\Delta$. For $\Delta>0$ an
evaluation of the pairing susceptibility with instantaneous pairing
interaction yields no Cooper instability\cite{Balatsky1993,Sudbo1995,Yin1996}.
Superconductivity would then only occur if the pairing interactions
exceeded a threshold value. Then a superconducting ground state would
be the exception rather than the rule. However, for a number of systems near
a fermionic quantum critical point, ranging from composite-fermion
metals, high-density quark matter to metals with magnetic or nematic critical
points, the self-consistently determined pairing interaction inherits
a singular behavior 
\begin{equation}
V_{{\rm pair}}\left(\omega\right)=V_{0}\left|\omega\right|^{1-4\Delta}
\end{equation}
with the same exponent $\Delta$\cite{Bonesteel1996,Son1999,Abanov2001,Abanov2001b,Roussev2001,Chubukov2005,She2009,Moon2010,Metlitski2015,Raghu2015,Lederer2015,Wu2019}.
The singular pairing interaction compensates for the weakened ability
of Non-Fermi liquid electrons to form Cooper pairs. One obtains a
generalized Cooper instability and superconductivity for infinitesimal
$V_{0}$.
A particularly dramatic phenomenon is the pairing of fully
incoherent Non-Fermi liquid states, e.g. systems with a flat and structureless
spectral function   
\begin{equation}
A_{{\rm IC}}\left(\omega\right)=A_{0} + \cdots.
\label{eq: incoh spectral f}
\end{equation}
The pairing of such fully incoherent fermions remains an open question.
It corresponds to the extreme limit of $\Delta=\frac{1}{2}$ of the
quantum-critical pairing problem. 

Significant progress in our understanding of quantum-critical superconductivity was achieved because of advances to formulate models that allow for sign-problem free Quantum-Monte-Carlo simulations \cite{Berg2012,Schattner2016a,Schattner2016b,Dumitrescu2016,Lederer2017,Li2017,Wang2017MC,Esterlis2018b,Berg2019}. The appeal of these computational approaches is that they allow for a detailed analysis of the interplay between quantum criticality, pairing, and other competing states of matter. Advances have also been made in clearly specifying how one would sharply distinguish the pairing state of a Non-Fermi liquid from the more conventional one.
Cooper pairing of quantum critical fermions and incoherent pairing should  be discernible  by analyzing  the frequency and temperature dependence of the dynamical pair susceptibility\cite{Chubukov2005,She2009,She2011}, a quantity accessible through higher-order Josephson effects.

An interesting approach that yields Non-Fermi liquid behavior
is provided by the Sachdev-Ye-Kitaev (SYK) model\cite{Sachdev1993,Georges2000,Sachdev2010,Kitaev2015,Kitaev2015b}
and generalizations thereof\cite{Sachdev2015,Maldacena2016,Polchinski2016,Fu2017,Bi2017,Song2017,Chowdhury2018}.
The SYK model describes $N$ fermions with a random, infinite-ranged
interaction and gives rise to a critical phase where fermions have
a vanishing quasi-particle weight at low energies and temperatures.
The model is exactly solvable in the limit of infinitely many fermions,
$N\rightarrow\infty$, yielding a tractable example of strong-coupling,
Non-Fermi liquid behavior. The SYK model is appropriate for situations
where interactions dominate over the kinetic energy. Thus, it could
serve as a toy model for systems that are characterized by flat bands,
such as cuprate superconductors for momenta near the anti-nodal points
or possibly twisted bilayer graphene near the magic angle\cite{Cao2018}. Another
appeal of this model is that its gravity dual is an asymptotic
Anti-de-Sitter space AdS$_{2}$ that can be explicitly constructed\cite{Kitaev2015b,Maldacena2016}
an approach that is particularly promising if one wants to include
fluctuations that go beyond the leading large-$N$ limit\cite{Bagrets2016,Bagrets2017}.

An exciting question is whether one can construct superconducting
versions of the SYK model and address the question of how pairing
occurs in such a Non-Fermi liquid state of matter. Indeed, in Ref.\cite{Patel2018}
Patel \emph{et al.} added an additional pairing interaction to the
model and demonstrated that an instantaneous attractive coupling induces
a large superconducting gap in the spectrum. This describes the behavior
of a Non-Fermi liquid towards Cooper pairing due to an interaction
that is unrelated to the initial cause of Non-Fermi liquid behavior.
In another setting, of neutral fermions coupled to a single
site of an ``ordinary'' complex spinless fermion, odd-frequency
superconductivity was recently discussed in Ref.\cite{Gnezdilov2019}. It was also shown 
recently by Y. Wang in Ref.~\cite{Wang2019} that superconductivity can emerge at $O(1/N)$ 
in a model similar to that discussed here (but in which superconductivity is absent in the large-$N$ limit).

A fundamental question is to understand systems where the interaction that  causes of the breakdown of the quasiparticle description is equally responsible for pairing. Such quantum-critical pairing is
then directly linked to the Non-Fermi liquid state. As we will see, the SYK-strategy
allows to construct a solvable model of superconductivity
near a quantum-critical point. Such a model has the potential to deepen our understanding of holographic superconductivity\cite{Hartnoll2008,Hartnoll2008b,Hartnoll2018}. The SYK model offers an explicit gravity dual that will have to display  an instability due to the onset of superconductivity.   

In this paper we present a model of electrons interacting with phonons 
via a random, infinite-range coupling. It is well established
that singlet superconductivity can easily be destroyed if one breaks
time-reversal symmetry. Thus, we consider a distribution function
of real-valued electron-phonon coupling constants. This will indeed
give rise to superconductivity in the SYK-model at leading order in
an expansion for large number of fermions and bosons. The well-known Eliashberg equations of superconductivity\cite{Eliashberg1960,Scalapino1969,Carbotte1990},
yet with self-consistently determined electron and phonon propagator,
turn out to be exact. 

Our calculation reveals that superconductivity emerges very differently
in the weak and strong coupling regime of the system. At weak coupling
$T_{c}$ coincides, up to numerical prefactors, with the crossover
from Fermi liquid to Non-Fermi liquid behavior. Such behavior, where
superconductivity preempts the ultimate quantum-critical state, is
reminiscent of that observed in heavy-electron\cite{Mathur1998,Petrovic2001,Nakatsuji2008,Knebel2011} and iron-based\cite{Kasahara2010,Boehmer2014,Shibauchi2014,Kuo2016}
superconductors. Thus, the superconducting state masks large parts
of the Non-Fermi liquid regime. Similar behavior was recently seen in Quantum-Monte-Carlo simulations of spin-fluctuation-induced superconductivity\cite{Berg2019}.
 The nature of the superconductivity changes in the strong-coupling regime, where pairing occurs deep in the Non-Fermi liquid state and
$T_{c}$ approaches a universal value times the bare phonon frequency.
Pairing at strong coupling is a genuine example of Cooper pairs made
up of completely ill defined individual electrons, a phenomenon that is relevant for the underdoped cuprate superconductors.  A model for incoherent fermions in the cuprates due to similarly soft bosons, that also gives rise to magnetic precursors,  was discussed in Ref.\cite{Schmalian1998,Schmalian1999} and is similar in spirit to the behavior found here in the strong coupling regime.
The resulting phase diagram that follows from our analysis is given in Fig.~\ref{Fig: phase diagram schem}.

The results of this paper are determined from a model of electrons that interact strongly with soft lattice vibrations. In several instances we compare the qualitative features of our results with observations made in strongly-correlated superconductors such as members of the  heavy fermion, iron-based, or cuprate family. Strong evidence exists that the pairing mechanism in these systems is predominantly of electronic origin. The findings of our analysis can however be  rather straightfowardly extended to models of electrons that interact with collective electronic excitations, such as nematic   or magnetic fluctuations; see also the summary section of this paper. In this more general reasoning do we see the justification of our statements as they pertain to the mentioned materials. 

\section{The Model}

We start from the following Hamiltonian:
\begin{eqnarray}
H & = & -\sum_{i=1}^{N}\sum_{\sigma=\pm}\mu c_{i\sigma}^{\dagger}c_{i\sigma}+\frac{1}{2}\sum_{k=1}^{M}\left(\pi_{k}^{2}+\omega_{0}^{2}\phi_{k}^{2}\right)+\frac{\sqrt{2}}{N}\sum_{ij,\sigma}^{N}\sum_{k}^{M}g_{ij,k}c_{i\sigma}^{\dagger}c_{j\sigma}\phi_{k},\label{eq:Hamiltonian}
\end{eqnarray}
with fermionic operators $c_{i\sigma}$ and $c_{i\sigma}^{\dagger}$
that obey $\left[c_{i\sigma},c_{j\sigma'}^{\dagger}\right]_{+}=\delta_{ij}$$\delta_{\sigma\sigma'}$
and $\left[c_{i\sigma},c_{j\sigma}\right]_{+}=0$ with spin $\sigma=\pm1$.
In addition we have phonons, i.e. scalar bosonic degrees of freedom
$\phi_{k}$ with canonical momentum $\pi_{k}$, such that $\left[\phi_{k},\pi_{k'}\right]_{-}=i\delta_{kk'}$.
Here $i,j=1\cdots N$ refer to fermionic modes and $k=1\cdots M$
to the phonon field. Below we consider the limit $N=M\rightarrow\infty$.
We briefly comment on the behavior for arbitrary $M/N$ in Appendix C.
For simplicity we assume particle-hole symmetry which yields $\mu=0$
for the chemical potential. Notice, the coupling to phonons usually shifts the particle-hole symmetric point to non-zero value of $\mu$. This is a consequence of the Hartree diagram.  However, this contribution vanishes in the $N \rightarrow \infty$ limit.

The electron-phonon coupling constants $g_{ij,k}$ are real, Gaussian-distributed
random variables that obey 
\begin{equation}
g_{ij,k}=g_{ji,k}.\label{eq:herm}
\end{equation}
The distribution function has zero mean and a second moment $\overline{\left|g_{ij,k}\right|^{2}}=\bar{g}^{2}$. 
The unit of $\bar{g}$ is energy$^{3/2}$. Thus,  the model has two energy scales, the bare phonon frequency $\omega_{0}$
and $\bar{g}^{2/3}$. For convenience we measure all energies and
temperatures in units of $\omega_{0}$ and use the dimensionless coupling
constant $g^{2}=\bar{g}^{2}/\omega_{0}^{3}$. Whenever it seems useful,
we will reintroduce $\omega_{0}$ in the final results.

We perform the disorder average using the replica trick\cite{Edwards1975}.
Since $g_{ij,k}$ only occurs in the random part of the interaction
we are interested in the following average
\begin{equation}
\overline{e^{-S_{{\rm rdm}}}}=\overline{e^{-\sum_{ijk}g_{ijk}O_{ijk}}},
\end{equation}
where $O_{ijk}=\frac{\sqrt{2}}{N}\sum_{\sigma a}\int_{0}^{\beta}d\tau c_{i\sigma a}^{\dagger}\left(\tau\right)c_{j\sigma a}\left(\tau\right)\phi_{ka}\left(\tau\right).$
Here, $a=1,\cdots,n$ stands for the replica index and the over-bar
denotes disorder averages, while $\tau$ stands for the imaginary
time in the Matsubara formalism with $\beta=\left(k_{B}T\right)^{-1}$
the inverse temperature. The $g_{ij,k}$ are for given $k$ chosen
from the Gaussian orthogonal ensemble (GOE) of random matrices\cite{Mehta2004}.
We obtain for the disorder average 
\begin{equation}
\left.\overline{e^{-\sum_{ijk}g_{ijk}O_{ijk}}}\right|_{{\rm GOE}}=e^{\bar{g}^{2}\sum_{ijk}\left(O_{ijk}^{\dagger}+O_{ijk}\right)^{2}}.\label{eq:GOE}
\end{equation}
There is an important distinction between the models with and
without time-reversal symmetry for individual disorder configurations.
If we allow for complex coupling constants with $g_{ij,k}=g_{ji,k}^{*}$,
then, for given $k$, $g_{ij,k}$ would be chosen from the Gaussian
unitary ensemble (GUE). Performing the disorder average for the case
of the unitary ensemble yields
\begin{equation}
\left.\overline{e^{-\sum_{ijk}g_{ijk}O_{ijk}}}\right|_{{\rm GUE}}=e^{2\bar{g}^{2}\sum_{ijk}O_{ijk}^{\dagger}O_{ijk}}.\label{eq:GUE}
\end{equation}
As can be seen from the distinct behavior of the disorder averages
in Eq.\ref{eq:GUE} and \ref{eq:GOE}, the orthogonal ensemble with
time reversal symmetry contains, in addition to terms like $O_{ijk}^{\dagger}O_{ijk}$,
that also occur in the unitary ensemble, the anomalous terms $O_{ijk}^{\dagger}O_{ijk}^{\dagger}$
and $O_{ijk}O_{ijk}$. The anomalous terms can be analyzed at large $N$ by 
introducing anomalous propagators and self energies. These terms give rise
to superconductivity, see Appendix A.

The subsequent derivation of the self-consistency equations of the
model in the large-$N$ limit proceeds along the lines of other SYK
models\cite{Sachdev1993,Kitaev2015,Kitaev2015b,Maldacena2016,Polchinski2016,Sachdev2015,Patel2018,Gnezdilov2019}.
Assuming replica diagonal solutions, we obtain a coupled set of equations
for the fermionic and bosonic self energies and Green's functions.
This derivation is summarized in Appendix A. The most straightforward
formulation can be performed using the Nambu spinors $c_{i}=\left(c_{i\uparrow},c_{i\downarrow}^{\dagger}\right)$
in the singlet channel. Then we obtain the coupled set of equations
for the self energies:
\begin{eqnarray}
\hat{\Sigma}\left(\tau\right) & = & \bar{g}^{2}\tau_{3}\hat{G}\left(\tau\right)\tau_{3}D\left(\tau\right),\label{eq:GOE1}\\
\Pi\left(\tau\right) & = & -\bar{g}^{2}{\rm tr}\left(\tau_{3}\hat{G}\left(\tau\right)\tau_{3}\hat{G}\left(\tau\right)\right),\label{eq:GOE2}
\end{eqnarray}
with $D^{-1}\left(\nu_{n}\right)=\nu_{n}^{2}+\omega_{0}^{2}-\Pi\left(\nu_{n}\right)$
 and the fermionic Dyson equation in Nambu space $\hat{G}\left(\epsilon_{n}\right)^{-1}=i\epsilon_{n}\tau_{0}+\mu\tau_{3}-\hat{\Sigma}\left(\epsilon_{n}\right)$,
where $\tau_{\alpha}$ are the $2\times2$ Pauli matrices in Nambu
space. Here $\epsilon_{n}=\left(2n+1\right)\pi T$ and $\nu_{n}=2n\pi T$
are fermionic and bosonic Matsubara frequencies, respectively. These relations correspond to the Eliashberg equations of electron
phonon superconductivity, however with the inclusion of  the fully renormalized
 boson self energy. We use the standard parametrization
for $\hat{\Sigma}$ in Nambu space\cite{Eliashberg1960,Scalapino1969,Carbotte1990}:
\begin{equation}
\hat{\Sigma}\left(\epsilon_{n}\right)=\Sigma\left(\epsilon_{n}\right)\tau_{0}+\Phi\left(\epsilon_{n}\right)\tau_{1},
\end{equation}
where we dropped the terms proportional to $\tau_{3}$ and $\tau_{2}$ due to our
assumption of particle-hole symmetry and by fixing the
phase of the superconducting wave function, respectively. We will
also frequently use the parametrization 
\begin{equation}
\Sigma\left(\epsilon_{n}\right)=i\epsilon_{n}\left(1-Z\left(\epsilon_{n}\right)\right),
\end{equation}
where $Z\left(\epsilon_{n}\right)^{-1}$contains information about
the quasiparticle weight.

\section{Non-Fermi liquid Fixed Points in the normal state}

\begin{figure}
\includegraphics[scale=0.7]{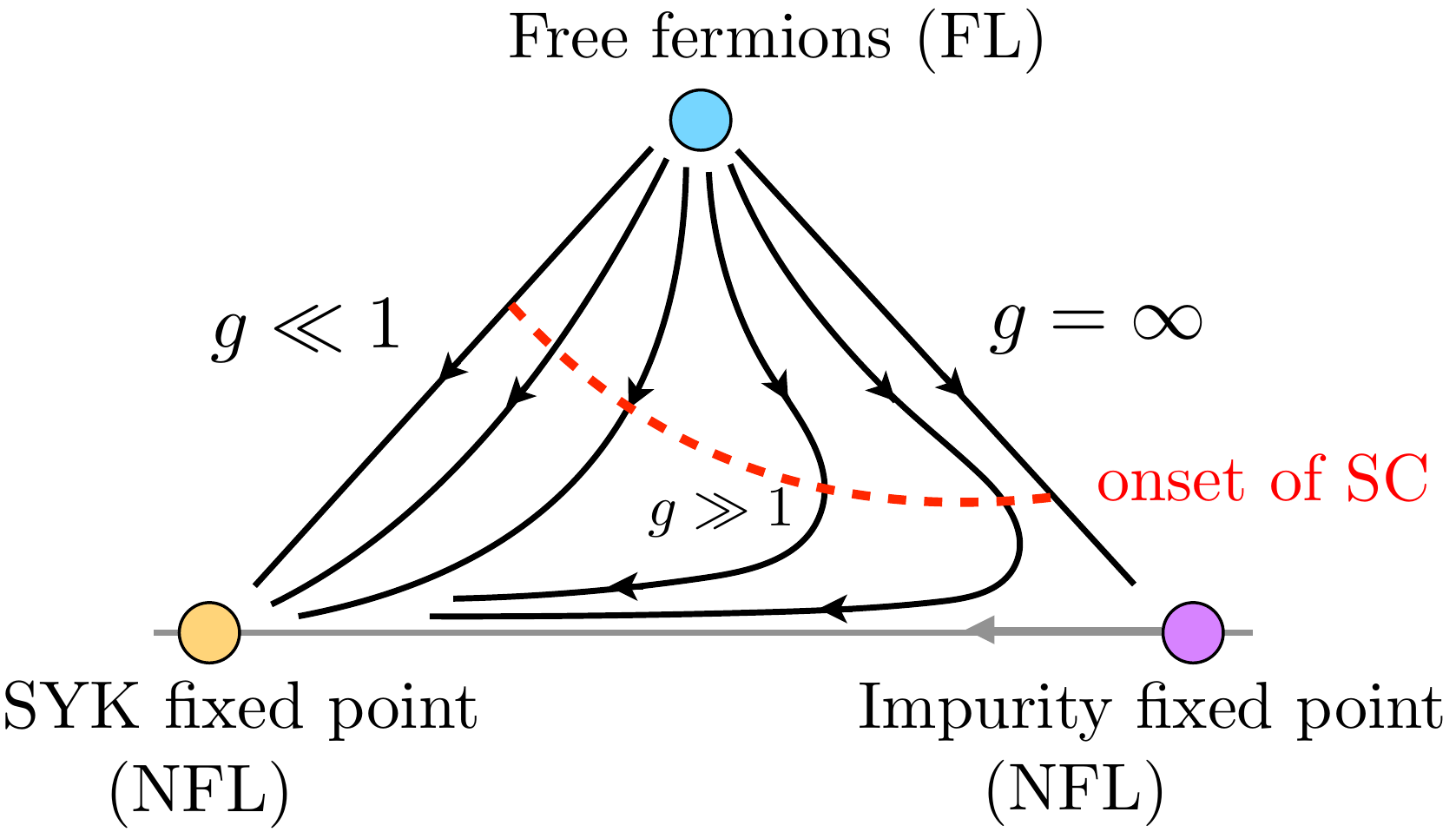}

\caption{Renormalization group flow that summarizes the physics of the phase
diagram of Fig.1. The free-fermion fixed point is always unstable
and flows at low energies to the quantum-critical SYK fixed point.
At strong coupling, the flow is influenced for a large energy window
by a new strong coupling fixed point of fully incoherent fermions.
At $g=\infty$ this impurity-like fixed point is stable and governs
the behavior at all scales. Superconductivity, marked by the red line,
at strong coupling occurs in the vicinity of the impurity-like fixed
point. At weak coupling it sets in at the crossover between the two
fixed points.}

\label{Fig: flow}
\end{figure}

We first solve the coupled equations in the normal state, i.e. assuming
that the anomalous self energy vanishes: $\Phi=0$. As discussed above
this corresponds to the full solution  of a model that
breaks time reversal symmetry for individual configurations of the
$g_{ij,k}$, chosen from the unitary ensemble. We obtain the following
coupled equations for the fermionic and bosonic self energies:
\begin{eqnarray}
\Sigma_{\sigma}\left(\tau\right) & = & \bar{g}^{2}G_{\sigma}\left(\tau\right)D_{0}\left(\tau\right), \label{eq:sc GUE1}\\
\Pi\left(\tau\right) & = & -\bar{g}^{2}\sum_{\sigma}G_{\sigma}\left(\tau\right)G_{\sigma}\left(-\tau\right),\label{eq:sc GUE2}
\end{eqnarray}
as well as the Dyson equations $G_{\sigma}^{-1}\left(\epsilon_{n}\right)=i\epsilon_{n}+\mu-\Sigma_{\sigma}\left(\epsilon_{n}\right)$
and $D^{-1}\left(\nu_{n}\right)=\nu_{n}^{2}+\omega_{0}^{2}-\Pi\left(\nu_{n}\right)$.
 As sketched in Fig.\ref{Fig: flow} these coupled equations give rise
to two distinct Non-Fermi liquid fixed points that govern the low
temperature regime for all coupling constants and the intermediate
temperature regime at strong coupling, respectively. In what follows
we will summarize the key properties of these fixed points, while
a detailed derivation of our results can be found in Appendix B.

\subsection{Low-temperature behavior: quantum-critical SYK-fixed point }

We first discuss the solution of this coupled set of equations at
low temperatures. The key finding is the following form of the fermionic
and bosonic propagators on the Matsubara axis:
\begin{eqnarray}
G\left(\epsilon_{n}\right) & = & \frac{1}{i\epsilon_{n}\left(1+c_{1}\left|\frac{g^{2}}{\epsilon_{n}}\right|^{2\Delta}\right)},\label{eq:QC1}\\
D\left(\nu_{n}\right) & = & \frac{1}{\nu_{n}^{2}+\omega_{r}^{2}+c_{3}\left|\frac{\nu_{n}}{g^{2}}\right|^{4\Delta-1}},\label{eq:QC2}
\end{eqnarray}
Here 
\begin{equation}
\omega_{r}^{2}=c_{2}\left(T/g^{2}\right)^{4\Delta-1}\label{eq:omega_r}
\end{equation}
is the renormalized phonon frequency. The $c_{i}$ are numerical coefficients
of order unity. The value of the exponent $\Delta$ is generally confined
to the interval $\frac{1}{4}<\Delta<\frac{1}{2}$, and for our problem
we find 
\begin{equation}
\Delta\backsimeq0.420374134464041.\label{eq:Delta}
\end{equation}
In Appendix B we derive these results, demonstrate that they agree
very well with our numerical solution of Eqs.\ref{eq:sc GUE1} and
\ref{eq:sc GUE2}, and give analytic  and numeric expressions for the coefficients
$c_{i}\left(\Delta\right)$. With $\Delta$ of Eq.\ref{eq:Delta}
we find $c_{1}\approx1.154700$, $c_{2}\approx0.561228$, and
$c_{3}\approx0.709618$. 

\begin{figure}
\includegraphics[scale=0.6]{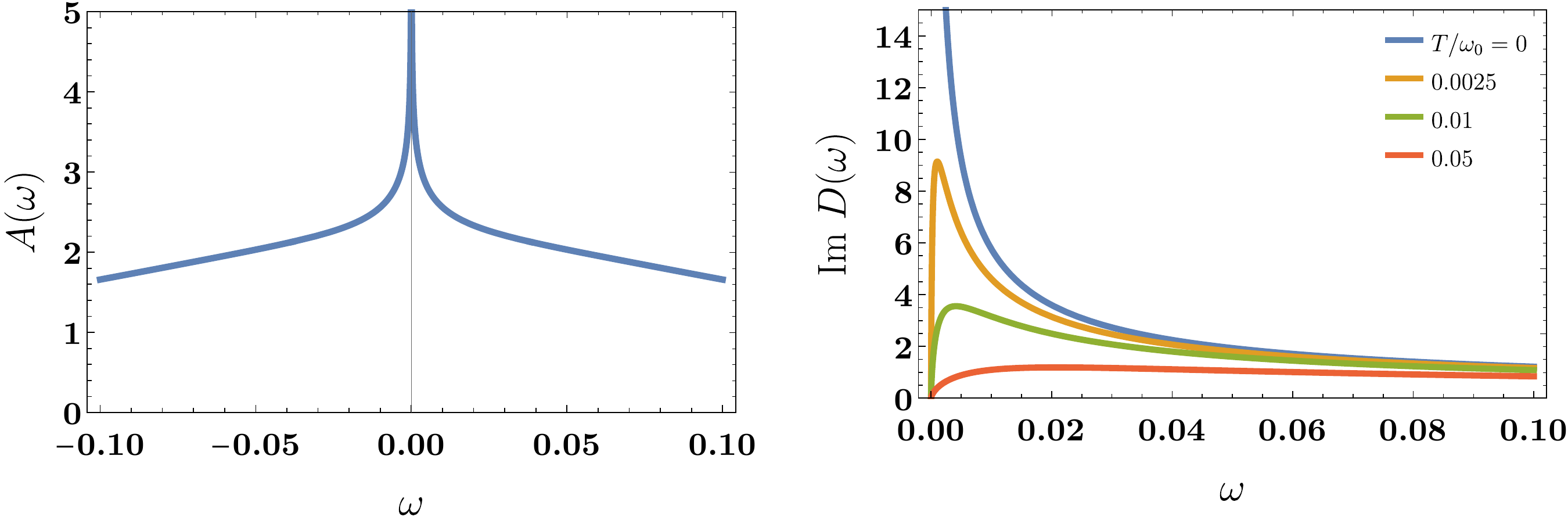}

\caption{Spectral function $A(\omega)=-\frac{1}{\pi} {\rm Im} G(\omega)$ and imaginary part of the bosonic propagator on
the real frequency axis for dimensionless coupling constant $g=0.5$.
The phonon spectrum is shown for several temperatures, displaying
the softening of the phonon mode $\omega_{r}$.}

\label{Fig_SYK_NFL_real_axis}
\end{figure}

The findings of Eqs.\ref{eq:QC1}, \ref{eq:QC2}, and \ref{eq:omega_r}
are summarized in Fig.\ref{Fig_SYK_NFL_real_axis}, where these equations have been
analytically continued from the imaginary to the real frequency axis. Let us discuss the main implications of these findings. The
fermionic propagator, Eq.\ref{eq:QC1} is similar to solutions of
other SYK models and at low energies is dominated by the self energy
\begin{equation}
\Sigma\left(\epsilon_{n}\right)=-i{\rm sign}\left(\epsilon_{n}\right)c_{1}g^{4\Delta}\left|\epsilon_{n}\right|^{1-2\Delta},
\end{equation}
with anomalous exponent $\Delta$. Only the numerical value of $\Delta$
is different from what can be found in purely fermionic models. Notice
however that we can vary $\Delta$ in the intervals $\left(\frac{1}{4},\frac{1}{2}\right)$
if we vary the ratio $M/N$ of the number of bosonic and fermionic
degrees of freedom, see Appendix C and Ref.\cite{Bi2017}. The bosonic
propagator Eq.\ref{eq:QC2} is, at low frequencies, dominated by an
anomalous Landau damping term, caused by the coupling to fermions
and hence determined by the same anomalous exponent $\Delta$. 

Notice that the system is critical for all values of $\omega_{0}$
and $g$. This is a surprising result. The renormalized phonon frequency
\begin{equation}
\omega_{r}^{2}=\omega_{0}^{2}-\Pi\left(0\right)
\end{equation}
in Eq.\ref{eq:omega_r} always vanishes as $T\rightarrow0$. One might
have expected that $\Pi\left(0\right)$ compensates the bare mass
only for one specific value of the coupling constant $g$, which would
then determine a quantum-critical point. Instead, the system remains
critical for all values of $g$, i.e. the fixed point described by
Eqs.\ref{eq:QC1} and \ref{eq:QC2} is stable, see Fig.\ref{Fig: flow}.
This stability is a consequence of the diverging charge susceptibility
of bare fermions with $G\left(i\epsilon_{n}\right)^{-1}\approx i\epsilon_{n}$.
It is the Non-Fermi liquid state that lifts the degeneracy of the
local Fermi liquid and protects the system against diverging charge
fluctuations. 

The scaling solution in Eqs.~\eqref{eq:QC1} and \eqref{eq:QC2} is valid in a low-temperature regime $T\lesssim T^*$ where the self-energies dominate the bare fermion and boson Green's functions. We can estimate this crossover temperature as 
\begin{equation}
T^* = \min(T_f, T_b),
\end{equation}
where $T_f \sim g^2\omega_0$ and $T_b \sim g^{-\phi}\omega_0$, where $0<\phi=\frac{8\Delta-2}{3-4\Delta} \leq 2$ for the allowed values $\frac{1}{4}<\Delta \leq \frac{1}{2}$. Below we will see that the relevant exponent at large $g$ is $\Delta=\frac{1}{2}$, so that $\phi=2$. Thus, the SYK-like quantum critical regime is confined to temperatures $T \lesssim g^2\omega_0$ at small $g$ and $T \lesssim g^{-2}\omega_0$ at large $g$ (see Fig.~\ref{Fig: phase diagram schem}).

%
%Finally it is worthwhile to estimate the regime of temperatures where
%these results apply. For $\left|\epsilon_{n}\right|\ll g^{2}$ the
%fermionic propagator is dominated by the anomalous power law. Thus,
%we expect the quantum critical SYK like behavior for temperatures
%below $T^{*}\propto g^{2}\omega_{0}$. However, the quantum critical
%behavior of the phonons is more restrictive if we consider the strong
%coupling limit $g\gg1$. The low frequency bosonic dynamics is governed
%by the fractional power for $\left|\nu_{n}\right|\ll g^{-\phi}$ where
%$0<\phi=\frac{8\Delta-2}{3-4\Delta}<2$ for the allowed values $\frac{1}{4}<\Delta<\frac{1}{2}$.
%Below we will see that the relevant exponent at large $g$ is $\Delta=\frac{1}{2}$
%such that $\phi=2$. Thus, the quantum critical regime Eqs.\ref{eq:GOE1}
%and \ref{eq:QC2} is, for large $g$, confined to frequencies and temperatures
%below $g^{-2}$, see Fig.~\ref{Fig: phase diagram schem}. 

\subsection{Intermediate-temperature behavior: impurity-like Non-Fermi liquid
fixed point}

The quantum critical regime of Eqs.\ref{eq:QC1} and \ref{eq:QC2} is
however not the only universal Non-Fermi liquid regime of this model.
Once $g>1$ an increasingly wide intermediate temperature window $g^{-2}<T<g^{2}$
opens up. In this new temperature window we find for the electron
and phonon propagators the solution
\begin{eqnarray}
G\left(\epsilon_{n}\right) & = & \frac{-2i{\rm sign}\left(\epsilon_{n}\right)}{\sqrt{\epsilon_{n}^{2}+\Omega_{0}^{2}}+\left|\epsilon_{n}\right|},\label{eq:SQC1}\\
D\left(\nu_{n}\right) & = & \frac{1}{\nu_{n}^{2}+\omega_{r}^{2}},\label{eq:SQC2}
\end{eqnarray}
with a large fermionic energy scale $\Omega_{0}=\frac{16}{3\pi}g^{2}$
and small phonon energy 
\begin{equation}
\omega_{r}^{2}=\left(\frac{3\pi}{8}\right)^{2}T/g^{2}.\label{eq:Somega_r}
\end{equation}

\begin{figure}
\includegraphics[scale=0.6]{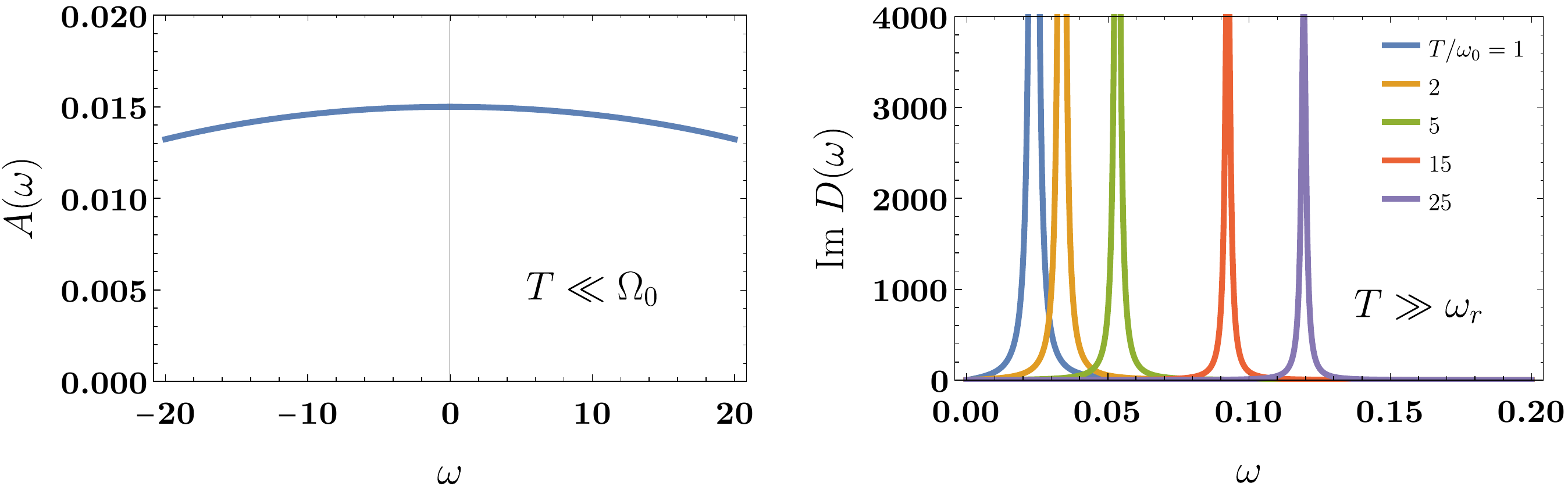}

\caption{Spectral function and imaginary part of the bosonic propagator on
the real frequency axis and for dimensionless coupling constant $g=5$.
The phonon spectrum is shown for several temperatures, displaying
the softening of the phonon mode $\omega_{r}$.}

\label{FIG_Imp_NFL_real_axis}
\end{figure}

The findings of Eqs.\ref{eq:SQC1}, \ref{eq:SQC2}, and \ref{eq:Somega_r}
are summarized in Fig.\ref{FIG_Imp_NFL_real_axis}. Since $T\ll\Omega_{0}$
fermions are ``cold'' and effectively behave as if they were quantum-critical with exponent $\Delta=\frac{1}{2}$, 
i.e. with impurity-like self energy 
\begin{equation}
\Sigma\left(\epsilon_{n}\right)=-i{\rm sign}\left(\epsilon_{n}\right)\frac{8}{3\pi}g^{2}.
\end{equation}
Non-interacting electrons with static impurities give rise to a similar self energy and can, for a given disorder configuration,  be  considered a Fermi liquid, essentially by definition. In our case the situation is different. We have to analyze multiple phonon configurations, even for a given disorder configuration of the $g_{ij,k}$. The resulting state cannot be mapped onto a free-fermion problem.  Hence the term Non-Fermi liquid.
The spectral function $A\left(\omega\right)$ is semicircular with
a width $2\Omega_{0}$. The low frequency spectral function is therefore
frequency independent
\begin{equation}
A\left(\left|\omega\right|\ll\Omega_{0}\right)=\frac{3}{8g^{2}},\label{eq: A impurity}
\end{equation}
reflecting the incoherent nature of the fermion spectrum, as mentioned in Eq.\ref{eq: incoh spectral f} in the introduction. On the other
hand, phonons are undamped but ``hot'', i.e. thermally excited since
$T\gg\omega_{r}$ once $T\gg g^{-2}$. Given the large fermionic
energy scale $\Omega_{0}$ we can neglect Landau damping terms that
we find to be $\propto\left|\omega_{n}\right|/\Omega_{0}$ in the
intermediate energy window. While the phonons are sharp excitations with a 
strongly renormalized, soft frequency, the fermions are highly incoherent.
Similar behavior was discussed in the context of magnetic precursors in cuprates\cite{Schmalian1998,Schmalian1999}.
The impurity-like behavior for the fermionic self energy is expected
given the quasi-static nature of the phonons. Notice, all these results
correspond to an anomalous fermionic exponent $\Delta=\frac{1}{2}$.
This strong-coupling fixed point is unstable and the system eventually
crosses over to the low-temperature SYK fixed point. Only for $g=\infty$ does 
the impurity fixed point describe the ultimate low-$T$ behavior,
see Fig.\ref{Fig: flow}. The analytic derivation of this strong coupling
criticality is summarized in Appendix B and compared with the full
numerical solution of Eqs.\ref{eq:sc GUE1} and \ref{eq:sc GUE2}.

\section{Superconductivity and Pairing of Non-Fermi Liquids}

In the previous section we analyzed the behavior of the model Eq.\ref{eq:Hamiltonian}
in the normal state. As indicated in Fig.\ref{Fig: phase diagram schem}
the normal state consists of three distinct regions that are separated
by crossover lines. Tor $T>T_{f}\approx g^{2}\omega_{0}$ interaction
effects are weak and we have essentially free electrons. For $T<T_{f}$
we have two distinct interacting regimes. At lowest temperatures with
$T<T^*\sim {\rm min}(g^{2}\omega_{0},g^{-2}\omega_{0}$) quantum-critical
behavior similar to that found in previous SYK-model calculations
occurs, where phonons are characterized by anomalous Landau damping.
For strong coupling, i.e. for $g>1$ a new universal intermediate
temperature window $g^{-2}<T/\omega_{0}<g^{2}$ opens up where strongly
incoherent fermions interact with soft phonons. 

\begin{figure}
\includegraphics[scale=0.8]{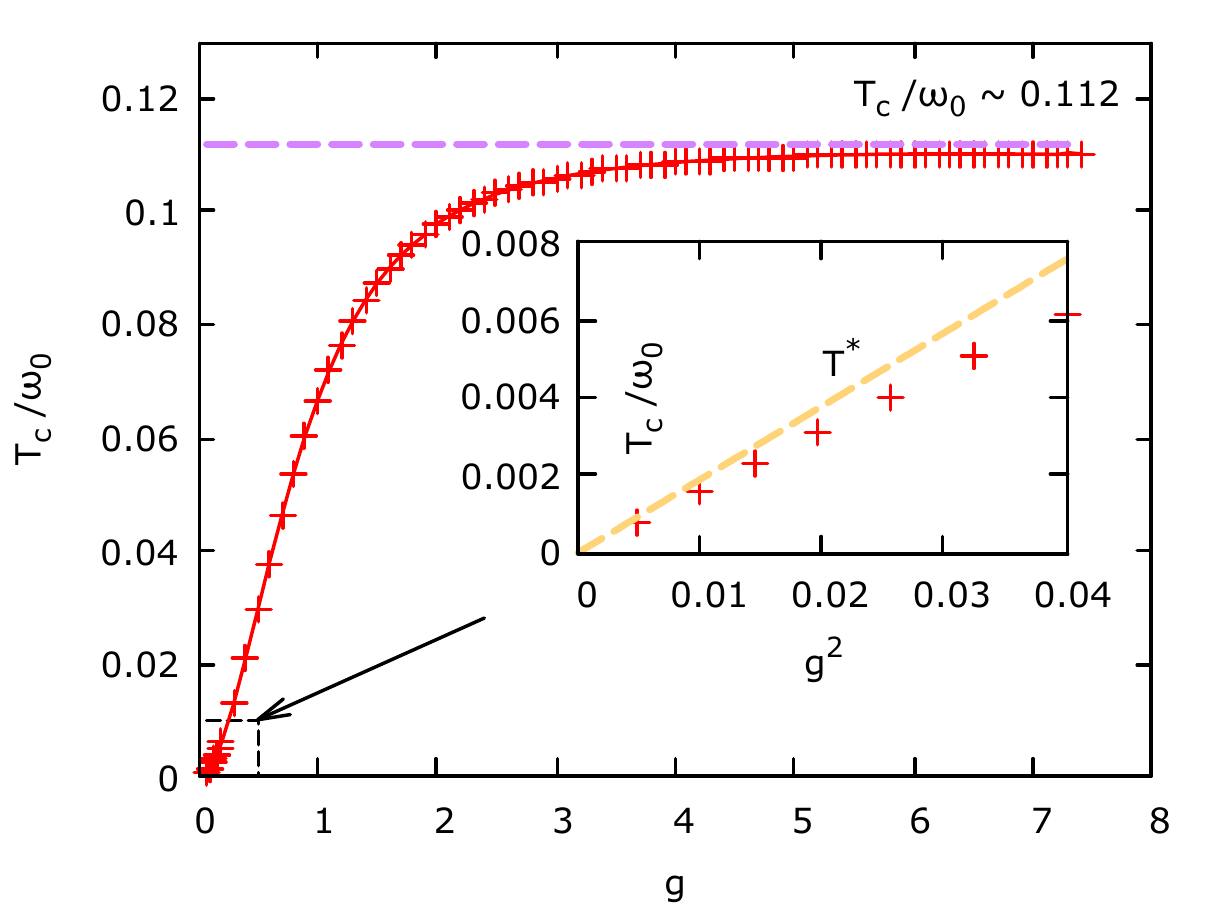}

\caption{Superconducting transition temperature as function of the coupling
constant from the numerical solution of the coupled equations in the
normal state and the analysis of the eigenvalue of the pairing vertex.
At weak coupling we obtain $T_{c}\propto g^{2}\omega_{0}$ while the
transition temperature saturates at strong coupling with $T_{c}\left(g\rightarrow\infty\right) \approx 0.112$$\omega_{0}$.}

\label{Fig: Tc numerical}
\end{figure}

Next we allow for superconducting solutions and solve the coupled
equations for the normal and anomalous self energies. On the Matsubara
axis, these coupled equations are
\begin{eqnarray}
i\epsilon_{n}\left(1-Z\left(\epsilon_{n}\right)\right) & = & -\bar{g}^{2}T\sum_{n^{\prime}}\frac{D\left(\epsilon_{n}-\epsilon_{n^{\prime}}\right)i\epsilon_{n'}Z\left(\epsilon_{n^{\prime}}\right)}{\left(\epsilon_{n'}Z\left(\epsilon_{n'}\right)\right)^{2}+\Phi\left(\epsilon_{n'}\right)^{2}},\nonumber \\
\Phi\left(\epsilon_{n}\right) & = & \bar{g}^{2}T\sum_{n'}\frac{D\left(\epsilon_{n}-\epsilon_{n^{\prime}}\right)\Phi\left(\epsilon_{n^{\prime}}\right)}{\left(\epsilon_{n'}Z\left(\epsilon_{n'}\right)\right)^{2}+\Phi\left(\epsilon_{n'}\right)^{2}} \nonumber \\
\Pi(\nu_n) &=& -2 \bar g^2 T \sum_{n'} [G(\epsilon_{n'} + \nu_n) G(\epsilon_{n'}) - F(\epsilon_{n'} + \nu_n) F(\epsilon_{n'})]
.\label{eq:Eliashb1}
\end{eqnarray}
If we linearize the second equation with respect to the anomalous
self energy $\Phi$ and set $\Phi=0$ in the first equation we can
determine the superconducting transition temperature. The result of
this analysis is summarized in Fig.\ref{Fig: Tc numerical}. First,
our model does indeed give rise to a superconducting ground state
for all values of the coupling constant $g>0$. For small $g$ the
transition temperature behaves as 
\begin{equation}
T_{c}\left(g\ll1\right)\approx 0.16g^{2}\omega_{0}.\label{eq:Tc weak coupl}
\end{equation}
Thus, while $T_{c}$ at weak coupling is numerically smaller than
the crossover scale $T^{*}$ to the quantum critical regime, both temperature
scales have the same parametric dependence. We will demonstrate in the next
section that indeed superconductivity at $g<1$ occurs near the onset
of the low-$T$ quantum critical state. The behavior changes at strong
coupling, where we find that 
\begin{equation}
T_{c}\left(g\rightarrow\infty\right)\approx0.11188\omega_{0}
\label{eq:Tc strong c}
\end{equation}
approaches a finite value. In this case we form Cooper pairs deep
in the Non-Fermi liquid state. We will analyze the behavior of this
new superconducting ground state and demonstrate that it is characterized
by a subtle formation of bound states of Cooper pairs with the dynamical
pairing field.

For our subsequent discussion it is useful to express the pairing
state in terms of the gap function 
\begin{equation}
\Delta\left(\epsilon_{n}\right)=\Phi\left(\epsilon_{n}\right)/Z\left(\epsilon_{n}\right).
\end{equation}
This yields the following coupled equations that are formally equivalent to
Eq.\ref{eq:Eliashb1}:
\begin{eqnarray}
Z\left(\epsilon_{n}\right) & = & 1+\bar{g}^{2}T\sum_{n^{\prime}}\frac{D\left(\epsilon_{n}-\epsilon_{n^{\prime}}\right)}{\sqrt{\epsilon_{n'}^{2}+\Delta^{2}\left(\epsilon_{n'}\right)}}\left[\frac{1}{Z\left(\epsilon_{n'}\right)\sqrt{\epsilon_{n'}^{2}+\Delta^{2}\left(\epsilon_{n'}\right)}}\right]\frac{\epsilon_{n^{\prime}}}{\epsilon_{n}},\nonumber \\
\Delta\left(\epsilon_{n}\right) & = & \bar{g}^{2}T\sum_{n'}\frac{D\left(\epsilon_{n}-\epsilon_{n^{\prime}}\right)}{\sqrt{\epsilon_{n'}^{2}+\Delta^{2}\left(\epsilon_{n'}\right)}}\left[\frac{1}{Z\left(\epsilon_{n'}\right)\sqrt{\epsilon_{n'}^{2}+\Delta^{2}\left(\epsilon_{n'}\right)}}\right]\left(\Delta\left(\epsilon_{n'}\right)-\frac{\epsilon_{n^{\prime}}}{\epsilon_{n}}\Delta\left(\epsilon_{n}\right)\right),\label{eq:Eliashberg2}
\end{eqnarray}
and the same equation for $\Pi(\nu_n)$. These equations are distinct from the usual Eliashberg theory where
the momentum integration over states in a broad band replaces the
terms in square brackets by $\pi\rho_{0}$, where $\rho_{0}$ is the density of states
in the normal state. In our problem we analyze systems
with non-dispering bands, changing the character of the Eliashberg equations.
We will see below that for very large $g$ the interactions give rise
to a significant broadening of the spectral function that allows to
replace the terms in square brackets by a spectral function $A\left(g\rightarrow\infty,\omega\right)=\frac{3}{8}g^{-2}$ times
$\pi$. In this limit some known results of the conventional Eliashberg
theory\cite{Carbotte1990,Allen1975,Marsiglio1991,Karakozov1991,Combescot1995}
can be used to obtain a better understanding of the strong coupling
limit.

The appeal of the reformulation in terms of the gap function in Eq.\ref{eq:Eliashberg2}
is that it clearly reveals the role of the zeroth bosonic Matsubara
frequency for the gap equation. Suppose the bosonic propagator is
dominated by the zeroth Matsubara frequency. This is the case at strong
coupling where we obtained with Eqs.\ref{eq:SQC2} and \ref{eq:Somega_r}
that $D\left(\nu_{m}\right)$ is dominated by $\nu_{m}=0$, a result
that led to the solutions of Eq.\ref{eq:SQC1}. From Eq.\ref{eq:Eliashberg2} it 
follows that there is no contribution to the pairing problem for $\epsilon_{n}=\epsilon_{n'}$.
Thus, static phonons do not affect the onset of superconductivity.
The same effect is also responsible for the Anderson theorem\cite{Anderson1959,Abrikosov1958a,Abrikosov1958b,Abrikosov1961,Potter2011,Kang2016}
where static non-magnetic impurities will not affect the superconducting
transition temperature. Soft phonons behave somewhat similar to non-magnetic
impurities\cite{Millis1988,Abanov2008}. Superconductivity is then only caused by the remaining
quantum fluctuations of the phonons. How this happens and what the
implications for the spectral properties of the superconducting state
are will be discussed in the subsequent sections.

\subsection{Superconductivity at weak coupling}

We start our analysis of superconductivity in the weak coupling regime
$g<1$ and first estimate the superconducting transition temperature
$T_{c}$ from the linearized version of Eq.\ref{eq:Eliashb1}
\begin{equation}
\Delta\left(\epsilon_{n}\right)=\bar{g}^{2}T\sum_{n'}\frac{D\left(\epsilon_{n}-\epsilon_{n^{\prime}}\right)}{Z\left(\epsilon_{n'}\right)\epsilon_{n'}^{2}}\left(\Delta\left(\epsilon_{n'}\right)-\frac{\epsilon_{n^{\prime}}}{\epsilon_{n}}\Delta\left(\epsilon_{n}\right)\right),
\end{equation}
where both $Z\left(\epsilon_{n}\right)$ and $D\left(\nu_{n}\right)$
are determined by our norml state solutions Eq.\ref{eq:QC1} and Eq.\ref{eq:QC2}.
Here we use $\epsilon_{n}Z\left(\epsilon_{n}\right)=\epsilon_{n}+i\Sigma\left(\epsilon_{n}\right)$.
For the phonon propagator of Eq.\ref{eq:QC2} we can safely neglect
the $\nu_{n}^{2}$ term in the denominator. When we explicitly write
out the temperature dependence in the various terms we obtain the
linearized gap equation 
\[
\Delta\left(\epsilon_{n}\right)=a_{0}\sum_{n'}\frac{\left(\frac{T_f}{T}\right)^{2\Delta}{\rm sign}\left(\epsilon_{n'}\right)}{\left(\frac{T}{T_f}\right)^{2\Delta}\left|n'+\frac{1}{2}\right|+\left|n'+\frac{1}{2}\right|^{1-2\Delta}}\frac{\frac{\Delta\left(\epsilon_{n'}\right)}{\epsilon_{n'}}-\frac{\Delta\left(\epsilon_{n}\right)}{\epsilon_{n}}}{m_{0}+\left|n-n'\right|^{4\Delta-1}},
\]
with $m_{0}=\frac{c_{2}}{c_{3}\left(2\pi\right)^{4\Delta-1}}\approx0.156558$,
$a_{0}=\frac{1}{2\pi c_{1}^{2}c_{2}}\approx0.212687$ and $T_f=\frac{1}{2\pi}c_{1}^{\frac{1}{2\Delta}}g^{2}\approx0.1888g^{2}$.
The temperature dependence of the gap equation only occurs in the
combination $T/T_f$. Thus the scale for the superconducting
transition is set by $T_f.$ However, this is precisely the
temperature scale where the crossover between the univeral low-$T$
non-Fermi liquid fixed point and the high-temperature free fermion
behavior takes place. This is also the reason why we included the
term $\left(\frac{T}{T_f}\right)^{2\Delta}\left|n'+\frac{1}{2}\right|$
in the denominator, which corresponds to the bare fermionic propagator.
Equally, the coefficient $m_{0}$ occurs as we have to include a finite
phonon frequency at the transition temperature. If we keep all those
terms we obtain $T_{c}\approx0.0821g^{2}$. Thus, we find that the
transition temperature is about half of the crossover temperature
$T_f.$ The $g^{2}$ dependence agrees with our numerical
finding shown in Fig.\ref{Fig: Tc numerical}. Not surprisingly, the
precise numerical coeffficient in $T_{c}$ cannot be reliably determined
as the transition temperature is right in the crossover regime between
free-fermion and quantum-critical SYK behavior. The reason is that there appear to be corrections to the fermionic 
self energy that are formally subleading at low frequencies, yet modify  numerical coefficients.
 The correct behavior was obtained from the full numerical solution and yields Eq.\ref{eq:Tc weak coupl}; see also Fig.\ref{Fig: Tc numerical}.

This analysis demonstrates that superconductivity in the weak coupling
regime occurs at the same temperature scale where quantum critical
Non-Fermi liquid behavior emerges. Thus superconductivity occurs instead
of the quantum critical regime. While parametrically the same, the
numerical coefficient of the transition temperature is somewhat smaller
than the crossover scale $T_f$ between the region of free
fermion and quantum-critical fermion behavior. Thus, in this regime
it might be possible to observe quantum critical scaling over a regime
up to a decade in frequency or temperature. It should however not
be possible to find several decades of universal scaling according
to Eqs.\ref{eq:QC1} and \ref{eq:QC2}. Superconductivity prevents
such a wide quantum-critical regime.

Nevertheless, it is very intructive to compare our gap function with
results from a previous analysis of the linearized gap-equation in
quantum-critical systems; see in particular Ref.\cite{Abanov2001,Chubukov2005,Moon2010,Metlitski2015,Raghu2015,Lederer2015,Wu2019}.
 If we formulate
the linearized gap equation merely in terms of the universal contributions
to the electron and phonon self energies, we obtain
\begin{equation}
\Phi\left(\epsilon_{n}\right)=\frac{T_{c}}{c_{1}^{2}c_{3}}\sum_{n'}\frac{\Phi\left(\epsilon_{n^{\prime}}\right)}{\left|\epsilon_{n}-\epsilon_{n'}\right|^{4\Delta-1}\left|\epsilon_{n'}\right|^{2-4\Delta}},
\label{eq:gapPhi}
\end{equation}
where $\epsilon_{n}=\left(2n+1\right)\pi T_{c}$. Here we can see
explicitly what was discussed in the introduction, namely that the
singular pairing interaction $V_{{\rm pair}}\left(\nu_{n}\right)\propto D\left(\nu_{n}\right)\propto \left| \nu_n \right|^{1-4\Delta}$
compensates for the less singular fermionic propagator giving rise
to a generalized Cooper instability. Self-consistency equations of
this type have been discussed in the context of several scenarios
for quantum critical pairing in metallic systems\cite{Bonesteel1996,Son1999,Abanov2001,Abanov2001b,Roussev2001,Chubukov2005,She2009,Moon2010,Metlitski2015,Raghu2015,Lederer2015,Wu2019}.
In this equation the entire $T$-dependence disappears given that
the two exponents in the denominator add up to unity. Thus, unless
this equation is supplemented by appropriate boundary conditions,
it is not possible to determine $T_{c}$, see Ref.\cite{Wu2019}. This is essentially achieved
by our above solution of the gap equation for $\Delta$$_{n}$. For a detailed discussion of the gap-equation in the form Eq.\ref{eq:gapPhi}, see Ref.\cite{Moon2010,Metlitski2015,Raghu2015,Lederer2015,Wu2019}.

\begin{figure}
\includegraphics[scale=0.8]{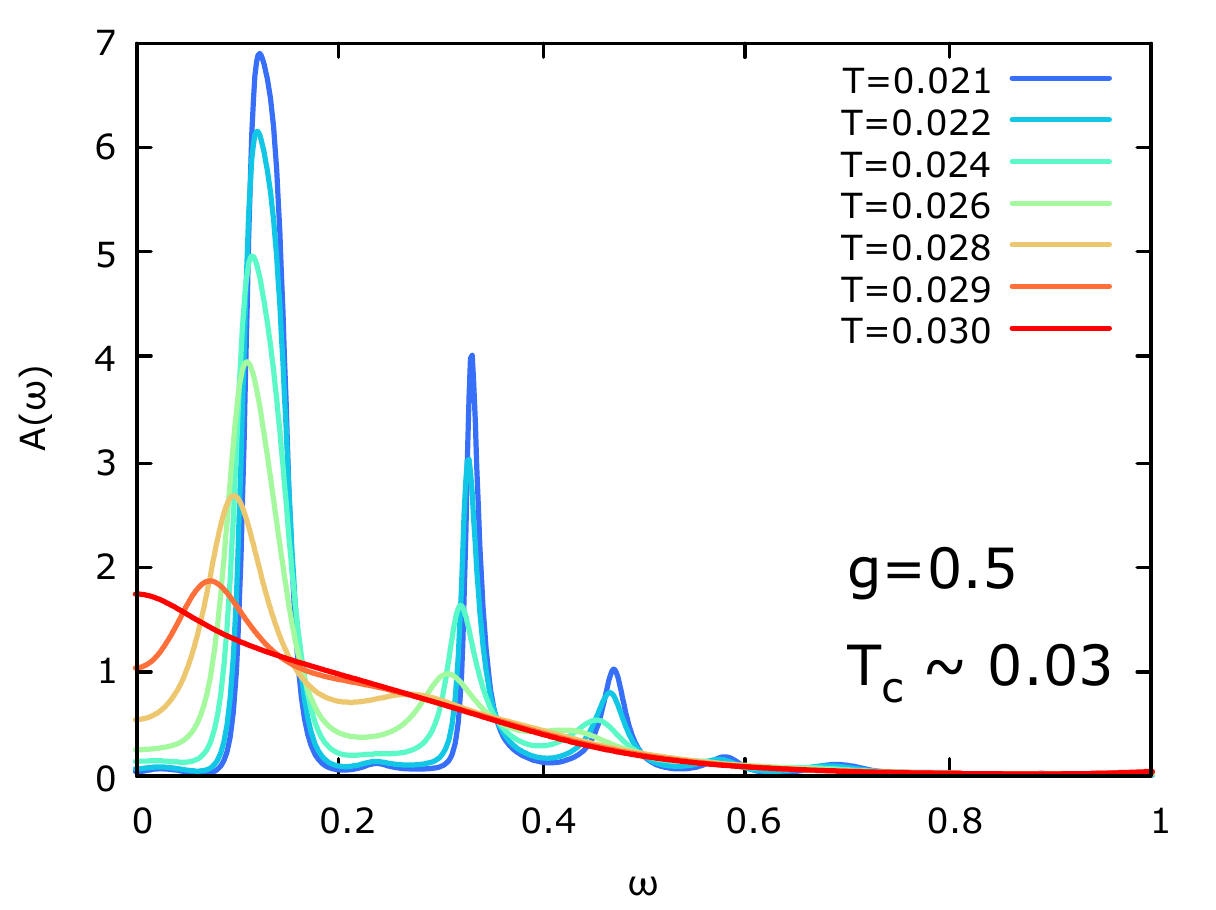}

\caption{Spectral function as function of temperature for $g=0.5$. The superconducting
transition temperature is $T_{c}=0.03\omega_{0}$. We find higher
order bound states as well as a gap closing as function of temperature.}

\label{Fig. SC spectral f weak c}
\end{figure}

In Fig.\ref{Fig. SC spectral f weak c} we show the spectral function
in the weak coupling regime at low temperatures that was obtained
from a numerical solution of the full coupled equations on the real
frequency axis, following the approach of Ref.\cite{Langer1995,CPC}.
Our main observation is the emergence of a sharp excitation,
and of several high energy structures. We will discuss these high
-energy shake-off peaks in greater detail in the discussion of the
strong coupling limit. Finally, we observe that in this weak coupling
regime the superconducting gap closes as the temperature increases.

Overall, the analysis of the pairing problem in this weak coupling
regime closely resembles the behavior that was found in a number of
metallic quantum critical points\cite{Bonesteel1996,Son1999,Abanov2001,Abanov2001b,Roussev2001,Chubukov2005,She2009,Moon2010,Metlitski2015,Raghu2015,Lederer2015,Wu2019}.
The SYK model proposed here may serve as a starting point to go beyond
the mean-field limit and investigate the fluctuation corrections by
following the advances in the $1/N$ corrections of SYK-like models\cite{Bagrets2016,Bagrets2017}.

\subsection{Superconductivity at strong coupling}

The investigation of superconductivity at strong coupling is of particular
interest, as it reveals why fully incoherent fermions are able to
nevertheless form a coherent superconducting state. We begin again
with a determination of the superconducting transition temperature
from the linearized gap equation. To this end we start from Eq.\ref{eq:Eliashberg2}
to obtain 

\begin{equation}
\Delta\left(\epsilon_{n}\right)=\frac{3\pi}{8}T_{c}\sum_{n'}\frac{1}{\left(\epsilon_{n}-\epsilon_{n'}\right)^{2}+\omega_{r}^{2}}\left(\frac{\Delta\left(\epsilon_{n'}\right)}{\epsilon_{n'}}-\frac{\Delta\left(\epsilon_{n}\right)}{\epsilon_{n}}\right){\rm sign}\left(\epsilon_{n'}\right).\label{eq:Eliashb3}
\end{equation}
Here, we used the normal state result Eq.\ref{eq:SQC1} that has the low frequency behavior
\begin{equation}
\left|\epsilon_{n}\right|Z\left(\epsilon_{n}\right)\approx\frac{8}{3\pi}g^{2}.
\end{equation}
The large normal state fermionic self energy is responsible for the
fact that the coupling constant $g$ gets cancelled in the prefactor
of Eq.\ref{eq:Eliashb3}. The only dependence on the coupling constant
in this equation is in the renormalized phonon frequency $\omega_{r}$.
At $T_{c}$, $\omega_{r}$ is determined by the normal state solution
of Eq.\ref{eq:Somega_r}. However, since $T\gg\omega_{r}$ in the
strong coupling regime and since the zeroth Matsubara frequency does
not contribute to superconductivity, we can simply set $\omega_{r}$
to zero in Eq.\ref{eq:Eliashb3}. The linearized gap equation becomes

\begin{equation}
\Delta_{n}=\alpha\sum_{n'\neq n}\frac{\frac{\Delta_{n'}}{2n'+1}-\frac{\Delta_{n}}{2n+1}}{\left(2n-2n'\right)^{2}}{\rm sign}\left(n'+\frac{1}{2}\right)\label{eq:Eliashb3-1}
\end{equation}
with $\alpha=\frac{3\omega_{0}^{2}}{8\pi^{2}T_{c}^{2}}$. One easily
finds that this equation has a solution for $\alpha_{c}=3.03458$,
which yields for the transition temperature $ T_{c}=\sqrt{\frac{3\omega_{0}^{2}}{8\pi^{2}\alpha_{c}}}\ $. Inserting the numerical coefficients yields Eq.\ref{eq:Tc strong c}.
The transition temperature saturates as $g\rightarrow\infty$, in quantitative agreement with
the numerical results shown in Fig.\ref{Fig: Tc numerical}. This
analysis also reveals the reason why pairing of fully incoherent fermions
is possible. The lack of fermionic coherence, with large imaginary part of
the electronic self energy, is caused by the coupling to almost static
bosonic modes. However, by arguments that in the context of disordered
superconductors give rise to the Anderson theorem, such static bosons
affect the normal and anomalous self energies $\Sigma$ and
$\Phi$, yet they cancel for the actual pairing gap $\Delta=\Phi/Z$
which is solely affected by the much weaker quantum fluctuations of
the bosonic spectrum. Thus, pairing of time-reversal partners occurs
even for incoherent fermions, a state that is protected by the same
mechanism that makes the superconducting transition temperature robust
against non-magnetic impurities\cite{Anderson1959,Abrikosov1958a,Abrikosov1958b,Abrikosov1961,Potter2011,Kang2016,Millis1988,Abanov2008}.

\begin{figure}
\includegraphics[scale=0.6]{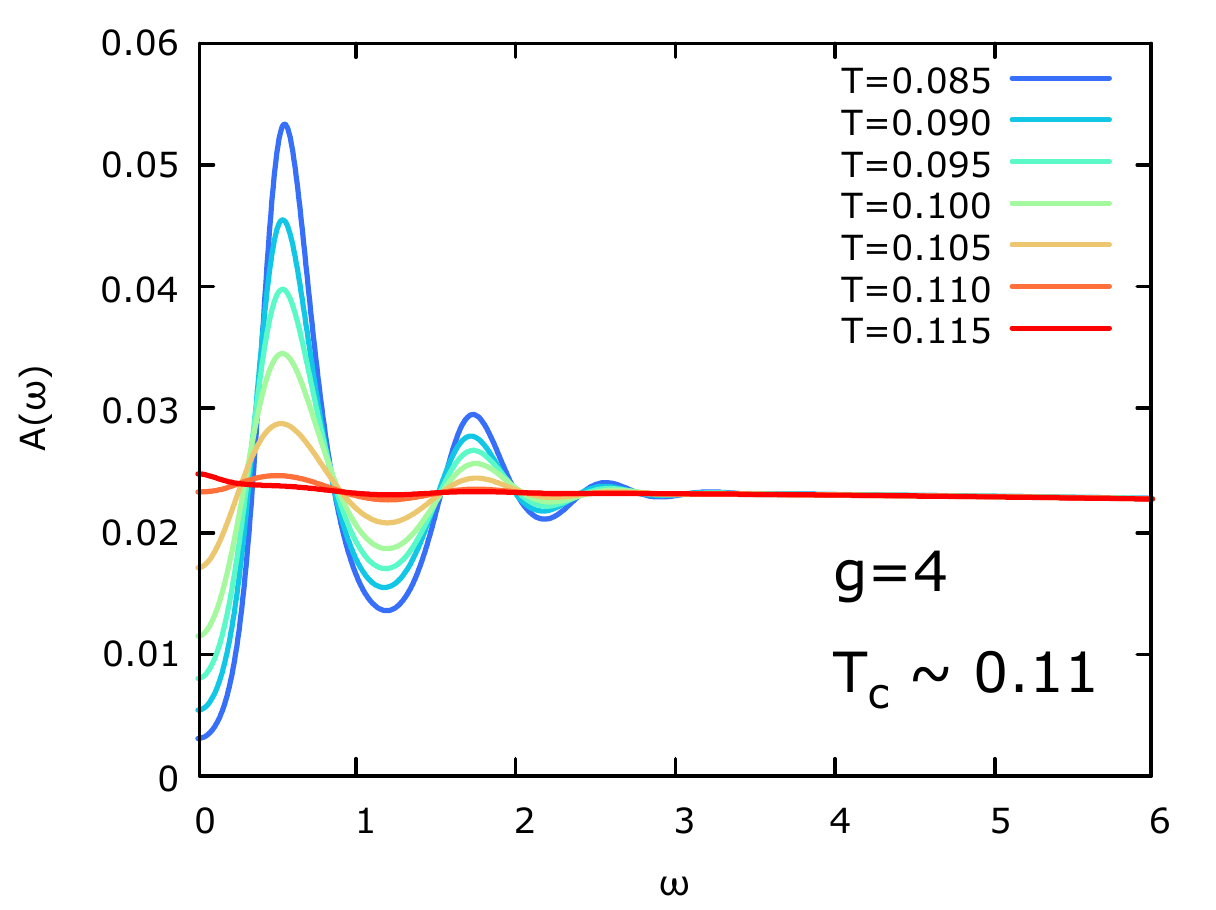}\includegraphics[scale=0.6]{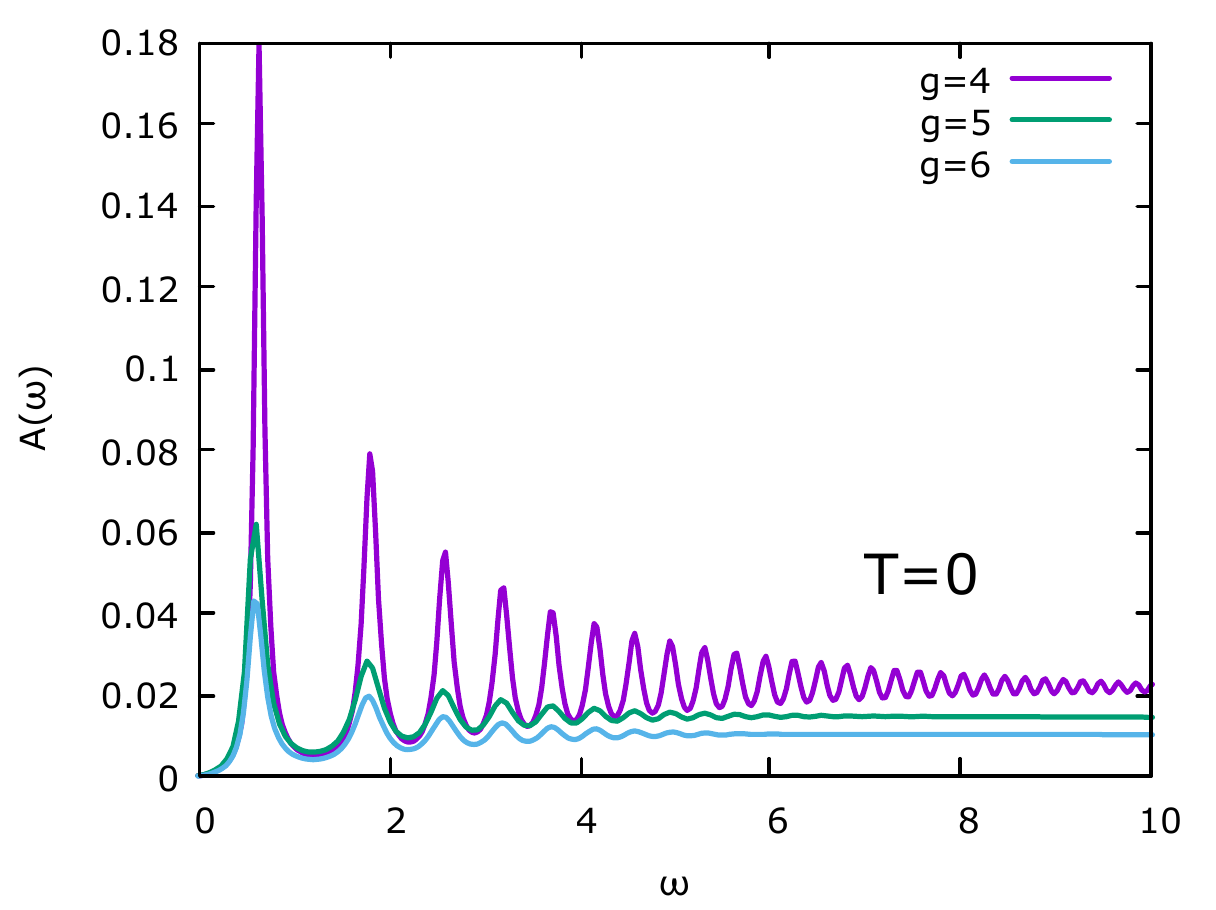}

\caption{Left panel: spectral function at strong coupling ($g=4$ with $T_{c}\approx0.11\omega_{0}$)
for different temperatures. In distinction to the weak coupling case
we find gap filling, rather than gap closing and a pronounced peak-dip-peak
structure. The latter is not due to the coupling to the phonon mode,
which has much smaller energy. Right panel: spectral function at $T=0$
for different coupling constants revealing a large number of shake-off
peaks that reflect the bound state formation in this limit of strongly
coupled Cooper pairs. Also, the total weight of the leading coherence
peak decreases with increasing coupling strength.}

\label{Fig: SC spectral function strong c}
\end{figure}

Now that we established that superconductivity sets in at a temperature
that is deep in the incoherent strong coupling regime, we discuss
the properties of this superconducting state. We start with our numerical
results for the spectral function and the anomalous Green's function.
In Fig.\ref{Fig: SC spectral function strong c} we show the fermionic
spectral function in the superconducting state. In distinction to
the gap-closing behavior that occurs at weak coupling, we find a filling
of the gap, where the position of the maximum is essentially unchanged with
temperature. In addition, higher order shake-off peaks occur that become
most evident in the strong coupling limit. The value of the superconducting
gap is, just like the transition temperature, independent of coupling
constant and of order of the bare phonon frequency $\omega_{0}$.
The lowest excitation of the fermions is $\Delta_0 \approx 0.640869140625\omega_0$. This yields 
\begin{equation}
2 \Delta_0/T_c\approx 11.456366,
\end{equation}
 which is more than three times the BCS value $2\pi e^{-\gamma_E}\approx 3.527754$. Such large values of  $2 \Delta_0/T_c$ have been obtained in the Eliashberg theory at strong coupling and for small phonon 
 frequencies\cite{Scalapino1969,Carbotte1990}; for a recent discussion see\cite{Wu2019b}.
 Since the spectral weight of the excited state is transferred
from energies below the gap, we can estimate the weight of the peak
as $Z_{{\rm coh}}\approx\int_{0}^{\Delta\approx\omega_{0}}A_{{\rm ns}}\left(\omega\right)d\omega\propto g^{-2}$,
where we used the normal state spectral function of Eq.\ref{eq: A impurity}.
We will see below that this result can be obtained rigorously at large
$g$. 

\begin{figure}
\includegraphics[scale=0.6]{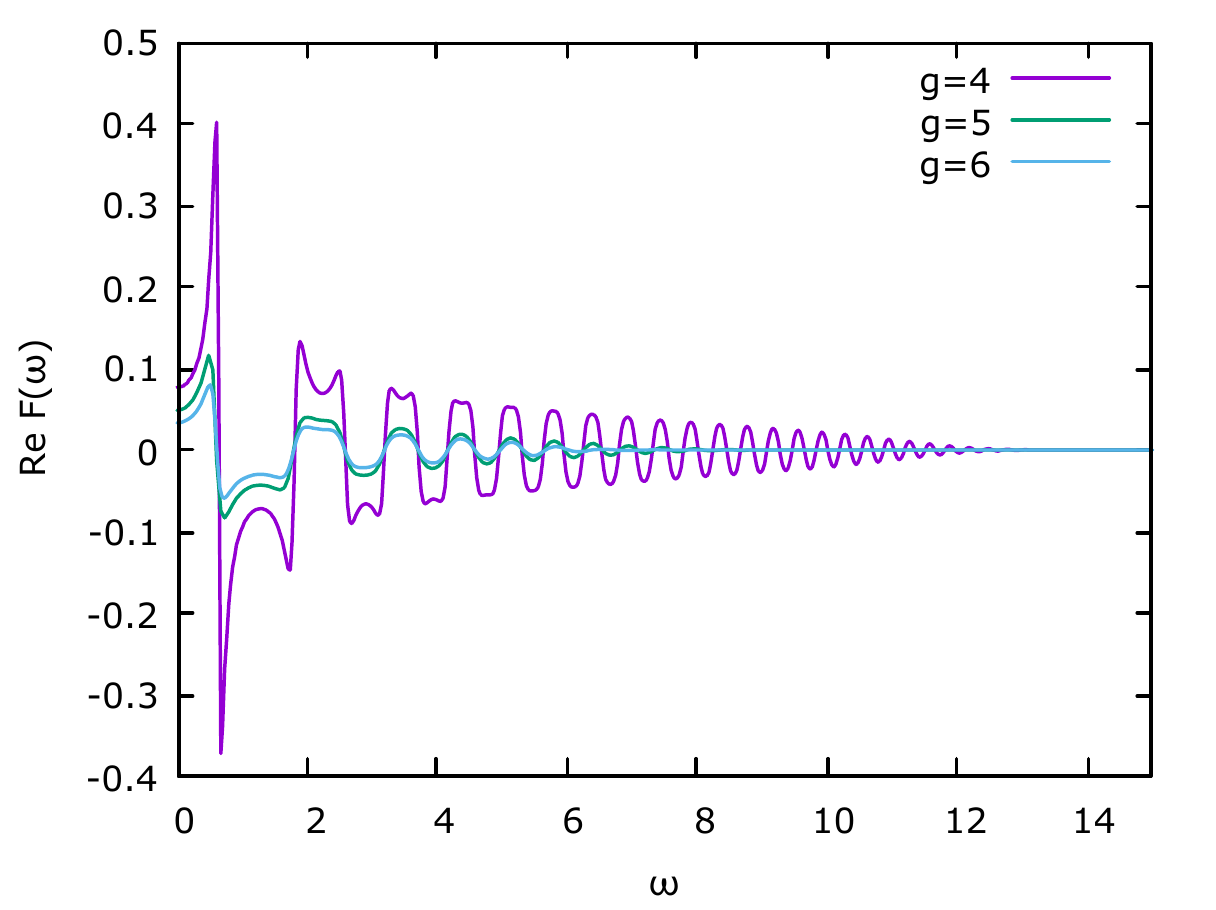}\includegraphics[scale=0.6]{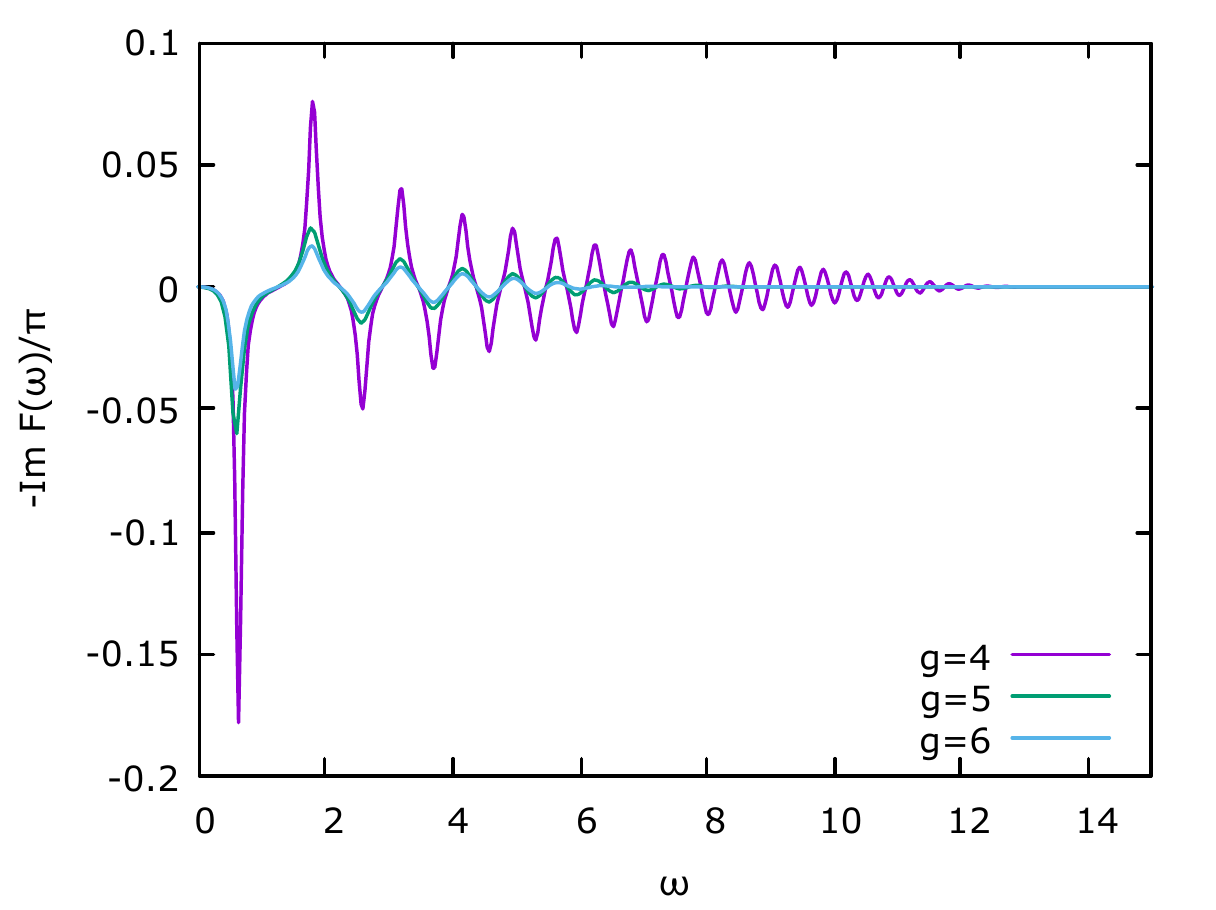}

\caption{Real part (left panel) and imaginary part (right panel) of the anomalous
propagator $F\left(\omega\right)$ at $T=0$ and for different coupling
strengths. Notice the alternating sign of the peaks in the imaginary
part.}

\label{Fig: SC anomalous GF strong c}

\end{figure}

A very intriguing feature of the low-$T$ spectral function is 
the occurrence of a large number of shake-off peaks at discrete energies $\Omega_l$ that are reminiscent
of the satellites that emerge as one forms polaronic states due to
strong electron-phonon coupling. However, in the conventional polaronic
theory these shake-off structures exist at energies $\epsilon_{0}+l\omega_{r}$
where $\epsilon_{0}$ is the bare fermion energy, $\omega_{r}$
the phonon frequency\cite{Mahan1993}, and $l$ an integer. In our case $\omega_{r}$ is
much smaller than the separation of the peaks in the spectral function.
In fact such structures in the normal and anomalous Greens function,
see Fig.\ref{Fig: SC anomalous GF strong c}, have already been discussed
in the context of strong coupling solutions of the Eliashberg theory\cite{Marsiglio1991,Karakozov1991,Combescot1995}
and can be considered as self trapping states of excited quasiparticles
in the pairing potential of the other electrons\cite{Combescot1995}. 
The excited quasiparticle polarizes the pairing field, that deforms and traps it. 
The positions of the peaks are not equidistant. Following Ref.\cite{Combescot1995} we find at large $l$ that the energies grow as $\Omega_l \approx \frac{\sqrt{3}\pi}{4}\sqrt{2l-1}\omega_0$. The first  ten peaks  are located at  $\Omega_l=p_l \Delta_0$ with $p_l=(1., 2.81, 4.05, 5.00, 5.76, 6.47, 7.14, 7.71, 8.29, 8.81)$. The first  peak corresponds of course to the  gap  $\Omega_1=\Delta_0$.
These features are a clear sign of the fact that we have strongly
interacting Cooper pairs, instead of an ideal gas of such pairs. While
most of these shake-off peaks smear out as the temperature increases 
(see left panel of Fig.\ref{Fig: SC spectral function strong c}) the
first one or two peaks should be visible and serve as potential
explanation for the observed peak-dip-hump structures seen in  photoemission spectroscopy measurements of cuprate superconductors near the antinodal momentum\cite{Dessau1991,ZXSchrieffer1997,Campuzano1996,Fedorov1999,Feng2000}.

\begin{figure}
\includegraphics[scale=0.8]{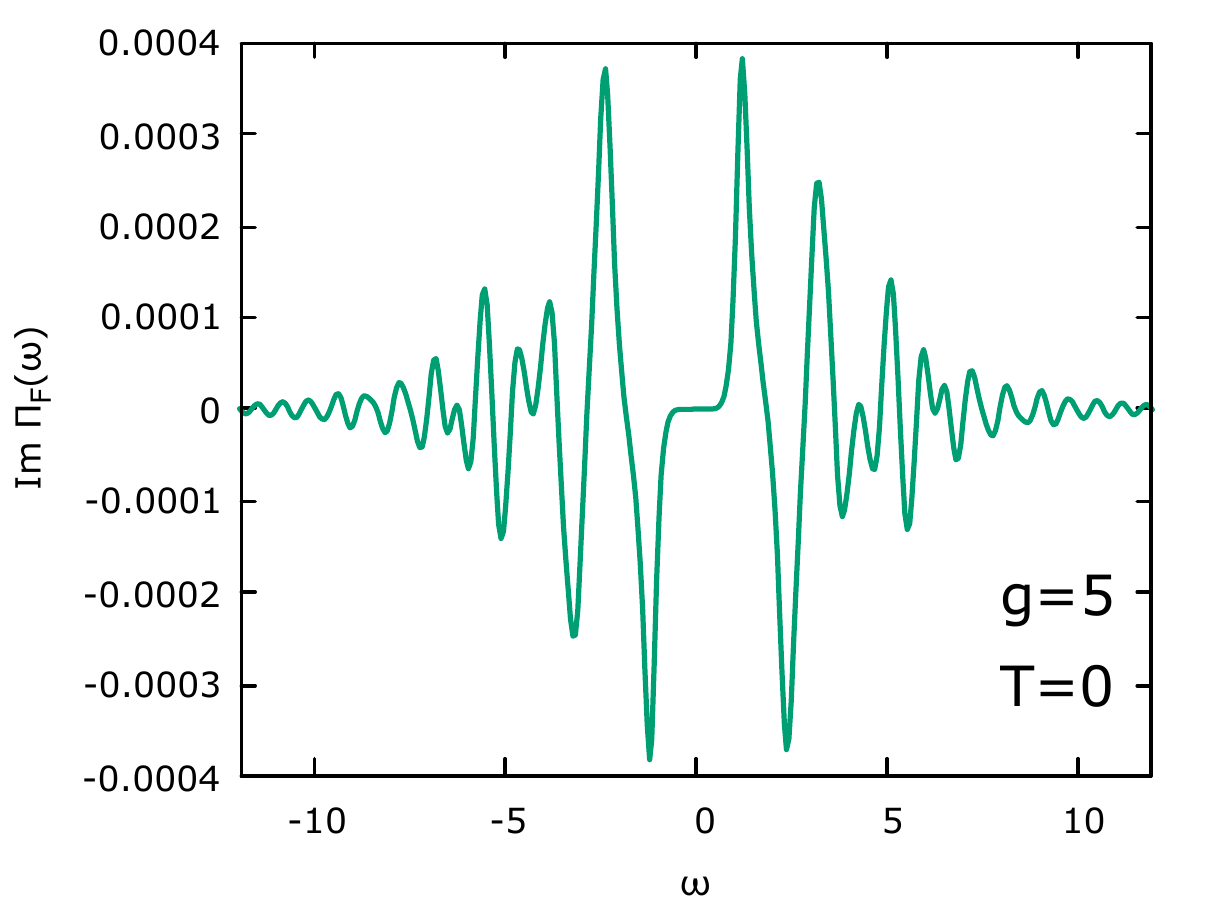}
\caption{Imaginary part of $\Pi_F(\omega)$ (defined in the text) for $g=5$ at $T=0$. ${\rm Im}\Pi_F(\omega)$ determines the amplitude of the phase-shifted AC Josephson current at higher voltage. The alternating sign of the peaks shown here is a direct consequence of the sign changes of consecutive peaks in the anomalous propagator, shown in Fig.\ref{Fig: SC anomalous GF strong c}. Thus, the AC-Josephson response might serve as a tool to identify the internal structure of the Cooper pair states of a strongly coupled superconductor.}
\label{Fig:ac_josephson}

\end{figure}

One way to verify the emergence of these shake-off peaks due to self
trapping in the pairing field is via the AC-Josephson effect with
current 
\begin{equation}
I_{J}\left(t\right)=2et_{0}^{2}\left({\rm Re}\Pi_{F}\left(eV\right)\sin\left(2eVt\right)+{\rm Im}\Pi_{F}\left(eV\right)\cos\left(2eVt\right)\right),
\end{equation}
where $\Pi_{F}\left(\omega\right)$ is the retarded version of the
Matsubara function $\Pi_{F}\left(\nu_{n}\right)=T\sum_{m}F^{\dagger}\left(\epsilon_{m}\right)F\left(\epsilon_{m}-\nu_{n}\right)$. 
At low applied voltage $\left|eV\right|<2 \Delta_0$ the imaginary part of $\Pi_F$ vanishes and the Josephson current is proportional to the sinus of the phase difference\cite{Josephson1962}. As the magnitude of the voltage exceeds $2\Delta_0$ an additional, phase-shifted AC Josepshon current that is proportional to  $\cos(2eVt)$ sets in \cite{Harris1974}. The coefficient is proportional to ${\rm Im}\Pi_{F}\left(eV\right)$ that we show in Fig.~\ref{Fig:ac_josephson}. Clearly the sequence of bound states of the spectral function can be identified in the cosine  AC-Josephson  response. Most interestingly, the sign change of two consecutive bound states, visible  in the anomalous propagator in Fig.\ref{Fig: SC anomalous GF strong c}, directly leads to an alternating sign of the phase-shifted Josephson signal. This offers a way to identify the nature of higher energy structures in the spectral function of superconductors, such as the bound states discussed here. For example, peaks in the spectral function due to multiple gaps on different Fermi surface sheets would not display such a sign-changing AC-Josephson signal.

\begin{figure}
\includegraphics[scale=0.8]{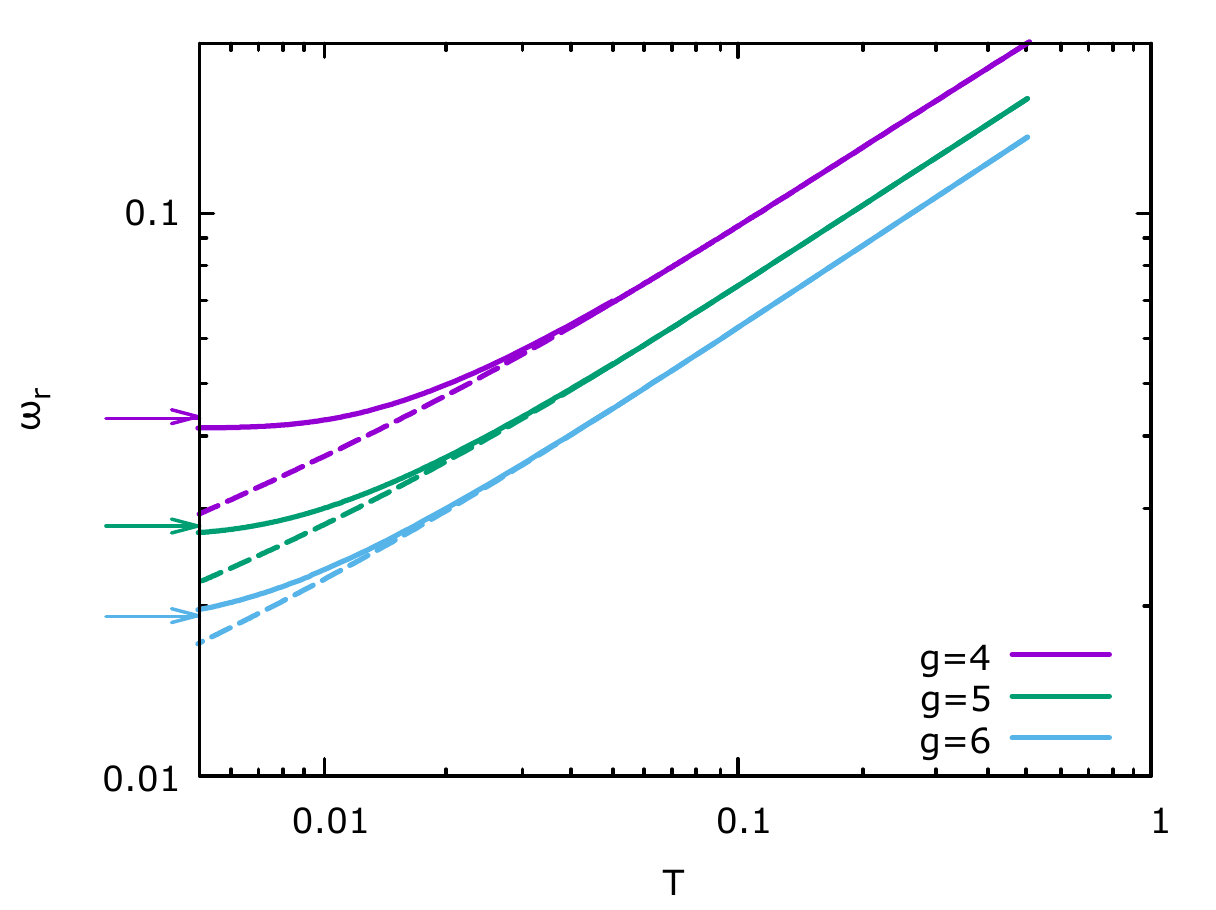}

\caption{Softening of the phonon frequency in the superconducting state at strong coupling. The dashed line is the normal state result, continued below $T_c$.
While in the normal state the phonon frequency vanishes as $T\rightarrow0$,
it approaches the finite $T=0$ value $\omega_{r}^{sc}=\frac{\omega_{0}}{2}\left(\frac{3\pi}{8}\right)^{2}g^{-2}$, indicated by the arrows. Thus, both the electrons and the bosons are gapped in the superconducting
state.}

\label{Fig. phononsoftening below Tc}
\end{figure}

Finally, in Fig.\ref{Fig. phononsoftening below Tc} we show our results
for the softening of the phonon frequency in the superconducting state.
In the normal state the phonon mode is expected to soften, first according
to Eq.\ref{eq:Somega_r} and below $T\sim\omega_{0}g^{-2}$ according to Eq.~\ref{eq:omega_r}. 
In the normal state $\omega_{r}$ always vanishes
for $T\rightarrow0$. With the onset of superconductivity the phonon
frequency still decreases with decreasing $T$, however it reaches
a finite value $\omega_{r}^{sc}$ at $T=0$. If we simply determine
the phonon renormalization from the high-energy behavior of the spectral
function in the superconducting state we find $\omega_{r}^{sc}=\frac{\omega_{0}}{2}\left(\frac{3\pi}{8}\right)^{2}g^{-2}$
which agrees well with our numerical finding. As expected the superconducting
ground state has gapped fermion and phonon excitations which explains
its coherent nature.

In the strong coupling limit one can make contact with results
that were obtained in the context of the usual Eliashberg theory,
where conduction electrons with a large bandwidth require momentum averaging\cite{Eliashberg1960,Scalapino1969,Carbotte1990}. This additional momentum
integration is not present in the SYK model, where one is interested
in the behavior of strongly-interacting narrow bands. From a purely technical point of
view, the effect of the momentum integration in the usual Eliashberg
formalism is to replace the term
\begin{equation}
{\cal A}\left(\epsilon_{n}\right)=\frac{1}{\pi}\frac{1}{Z\left(\epsilon_{n}\right)\sqrt{\epsilon_{n}^{2}+\Delta^{2}\left(\epsilon_{n}\right)}},\label{eq:calA}
\end{equation}
that occurs in square brackets in Eq.\ref{eq:Eliashberg2}, by the
normal state density of states of the system. We will
now show that at strong coupling the interaction-induced broadening
plays a similar role to the momentum integration and we can replace
${\cal A}\left(\epsilon_{n}\right)$ by the broad spectral function
of Eq.\ref{eq: A impurity}, i.e. ${\cal A}\left(\epsilon_{n}\right)\approx\frac{3}{8}g^{-2}$.
To demonstrate this we take the $T=0$ limit for $Z\left(\epsilon\right)$
in Eq.\ref{eq:Eliashberg2}:

\begin{equation}
Z\left(\epsilon\right)=1+\bar{g}^{2}\int\frac{d\epsilon'}{2\pi}\frac{1}{\left(\epsilon-\epsilon'\right)^{2}+\left(\omega_{r}^{sc}\right)^{2}}\frac{1}{Z\left(\epsilon'\right)\left(\epsilon'^{2}+\Delta^{2}\left(\epsilon'\right)\right)}\frac{\epsilon'}{\epsilon},
\end{equation}
At large $g$ the $T=0$ phonon frequency is small and the sharp Lorentzian
behaves as a $\delta$-function. Using our above result for $\omega_{r}^{sc}$ it 
follows that
\begin{equation}
Z\left(\epsilon\right)=1+\left(\frac{8g^{2}}{3\pi}\right)^{2}\frac{1}{Z\left(\epsilon\right)\left(\epsilon{}^{2}+\Delta^{2}\left(\epsilon\right)\right)},
\end{equation}
which yields at large $g$ the solution 
\begin{equation}
Z\left(\epsilon\right)=\frac{8g^{2}}{3\pi}\frac{1}{\sqrt{\epsilon{}^{2}+\Delta^{2}\left(\epsilon\right)}}.\label{eq:Z and Delta large g}
\end{equation}
Thus, while $Z\left(\epsilon\right)$ and $\Delta\left(\epsilon\right)$
depend strongly on frequency in the superconducting state, the combination
that enters ${\cal A}\left(\epsilon\right)$ is a constant. We have
verified that this result for $Z\left(\epsilon\right)$ agrees very
well with the full numerical solution for $g\gtrsim4$. Using Eq.\ref{eq:Z and Delta large g}
the equation for the gap function is given as 
\begin{equation}
\Delta\left(\epsilon_{n}\right)=\frac{3\pi}{8}T\sum_{n'}\frac{D\left(\epsilon_{n}-\epsilon_{n^{\prime}}\right)}{\sqrt{\epsilon_{n'}^{2}+\Delta^{2}\left(\epsilon_{n'}\right)}}\left(\Delta\left(\epsilon_{n'}\right)-\frac{\epsilon_{n^{\prime}}}{\epsilon_{n}}\Delta\left(\epsilon_{n}\right)\right).\label{eq: Delta at strong c}
\end{equation}
While the physics we are describing is rather different, formally
this equation is identical to the usual Eliashberg theory, yet with
a dimensionless coupling constant $\lambda=\frac{3}{8}$ and a very
soft phonon frequency. If we set this phonon frequency to zero, the
solution for $\Delta\left(\epsilon_{n}\right)$ is fully universal
and independent of the coupling constant. Comparing with the numerical
solution, we find that for $g\gtrsim4$ this is indeed the case with
high accuracy. Our result Eq.\ref{eq:Tc strong c} can also be obtained
from the well known strong coupling solution $T_{c}\approx0.1827$$\sqrt{\lambda}\omega_{0}$
by Allen and Dynes\cite{Allen1975} if one inserts $3/8$ for the
coupling constant. This is curious as one is very far from the applicability
of this strong-coupling Allen-Dynes result for $\lambda=0.375$. The
reason we can apply this formula is because of the extreme softening
of the phonons in our critical system. In the usual Eliashberg formalism
the frequency that enters the phonon propagator $D\left(\nu_{n}\right)$
is the bare, unrenormalized phonon frequency $\omega_{0}$. Then, the Allen Dynes limit of $T_c$ only becomes relevant for extremely large values of the couplig constant.

Using Eq.\ref{eq:Z and Delta large g} we can also find a very efficient
way to relate the function $\Delta\left(\omega\right)$ on the real
frequency axis and the spectral function 
\begin{equation}
A\left(\omega\right)=\frac{3}{8g^{2}}{\rm Re\left(\frac{\omega}{\sqrt{\omega^{2}-\Delta\left(\omega\right)^{2}}}\right)}.
\end{equation}
Since at large $g$ the solution for the gap function is independent
of the coupling constant, we immediately see that the weight of the
superconducting coherence peak must scale as $g^{-2}$, a behavior that
we estimated earlier based on the conservation of spectral weight.
Thus, the key effect of the incoherent nature of the normal state
is the reduced weight of the coherence peak, not its lifetime.

\begin{figure}
\includegraphics[scale=0.8]{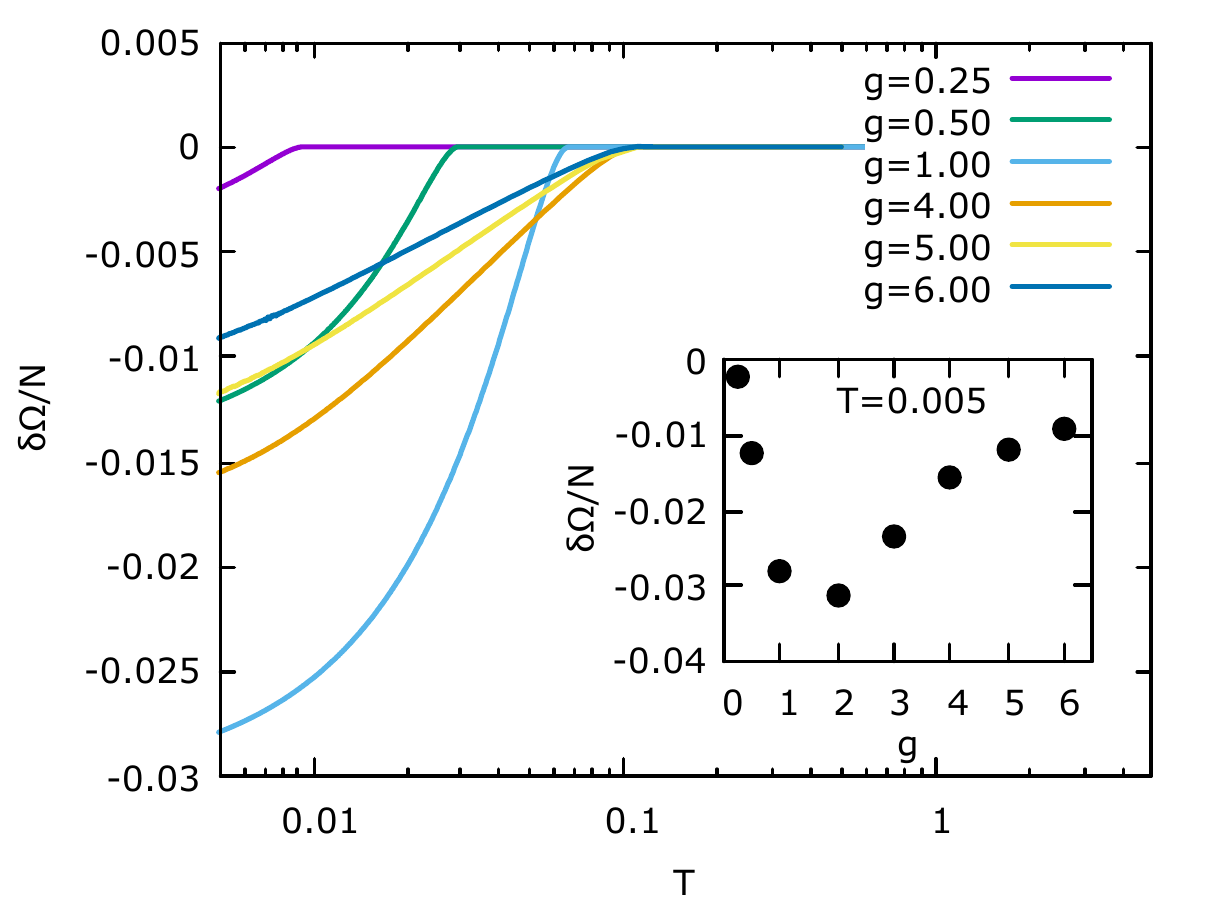}

\caption{Condensation energy $\delta \Omega/N$ as a function of temperature $T$ for several values of $g$. The inset shows $\delta \Omega/N$ as a function of $g$ at $T=0.005\omega_0$.}

\label{Fig:condensation energy}
\end{figure}

We finish this discussion with an analysis of the condensation energy
as function of temperature and coupling strength.  We determine the condensation energy $\delta \Omega$ from the difference of
\begin{eqnarray}
\Omega/N & = & -T\sum_{n}{\rm tr}\log\left(\hat{1}-\hat{G}_{0}\left(\nu_{n}\right)\hat{\Sigma}\left(\nu_{n}\right)\right)+\frac{T}{2}\sum_{m}\log\left(1-D_{0}\left(\epsilon_{m}\right)\Pi\left(\epsilon_{m}\right)\right),\nonumber \\
 & - & T \sum_{n}{\rm tr}\left(\hat{G}\left(\nu_{n}\right)\hat{\Sigma}\left(\nu_{n}\right)\right)
\end{eqnarray}
in the normal and superconductng state. Here, the trace is performed with respect to the degrees of freedom in Nambu space.
As shown in Fig.~\ref{Fig:condensation energy}, the temperature dependence of the condensation energy is very different in the weak and strong coupling regime with an almost linear behavior for large $g$. In this regime 
we  also find a close relation between the condensation energy and the quasiparticle
weight. At weak coupling $g<1$ the magnitude of the condensation energy rises precipitously with increasing $g$. On the other hand, for $g \gtrsim 4$ the magnitude of the condensation energy drops slowly, consistent with the power-law drop off of the quasiparticle weight.  Such a correlation between coherent weight in the superconducting state and condensation energy has indeed  been observed in the cuprate superconductors~\cite{Feng2000}.

\section{Summary}

In summary, we introduced and solved a model of interacting
electrons and phonons with random, infinite-ranged 
couplings that is in the class of Sachdev-Ye-Kitaev models and allows
for an exact solution in the limit of a large number of fermion and boson flavors.
The normal state phase diagram is summarized in Fig.~\ref{Fig: phase diagram schem}
and contains adjacent to a high energy regime of almost free fermions,
two distinct Non-Fermi liquid regimes. If the random electron-phonon
interaction respects time reversal symmetry not just on the average,
but for each disorder configuration, the system becomes superconducting.
Despite the incoherent nature of normal state excitations, sharp,
coherent excitations, including higher order shake-off peaks,
emerge below $T_{c}$. However, the broader the fermionic states above
$T_{c}$, the smaller the weight of the coherence peak below $T_{c}$.
The superconducting transition temperature grows monotonically with
coupling strength and levels off at a finite value that is determined
by the bare phonon frequency. We remark that 
a general upper bound on $T_c$ in conventional superconductors 
was recently proposed in Ref.~\cite{Esterlis2018}, with the numerical value 
$T_c \lesssim \bar\omega/10$ comparable to the maximal $T_c$ found here 
($\bar \omega$ is an appropriately defined maximal phonon frequency). However,
 in that case the bound is ultimately due to polaron physics at strong coupling, 
 which is absent in the $N \rightarrow \infty$ limit of the model considered here. 
 In contrast to $T_c$, we find the condensation energy
is non-monotonic and largest for intermediate coupling strength $g\approx1$.
Thus, we expect strong fluctuations for large $g$ if one goes beyond
the leading large-$N$ limit. Indeed, the appeal of the SYK formalism
is that it offers a well defined avenue to systematically improve
the results, see e.g. Refs.\cite{Bagrets2016,Bagrets2017}. Our analysis
can also be used as a starting point for lattice models of coupled
strongly-interacting superconductors and may be relevant in the theory
of Josephson-Junction arrays that are made up of unconventional superconductors.
Finally, our analysis was performed for fermions that interact with
a phonon mode, i.e. a scalar boson that couples to the fermion operator
$c_{i\sigma}^{\dagger}c_{j\sigma}$ in the charge channel. It is straightforward
 to generalize the model and include a spin-1 boson $\mathbf{\boldsymbol{\phi}}_{k}$
that couples to electrons via $g_{ij,k}\mathbf{\boldsymbol{\phi}}_{k}\cdot\sum_{\sigma\sigma'}c_{i\sigma}^{\dagger}\boldsymbol{\sigma}_{\sigma\sigma'}d_{j\mathbf{\sigma'}}$
with $\boldsymbol{\sigma}$ the vector of Pauli matrices in spin space
and with two fermion species $c_{i\sigma}$ and $d_{j\sigma}$. These two fermion species correspond to different bands or different antinodal regions  on the same band, depending on the problem under consideration.
The large-$N$ equations of this model are very similar to Eqn.\ref{eq:GOE1}
and \ref{eq:GOE2}, with $\tau_{3}\rightarrow\tau_{0}$. The superconducting
gap function of the two fermion species  then has a relative minus sign,
just like the gap function at the two anti-nodal points of a $d$-wave
superconductor. The formal expression for the gap function turns out to be the same as the one discussed in this paper.
 Overall, the approach presented here is a promising starting point to understand superconductivity in strongly coupled, incoherent materials. It justifies some of the known results of the Eliashberg formalism, in particular in the strong-coupling limit, and serves as a starting point to include fluctuations that go beyond the Eliashberg theory.

\emph{Note Added:} After the completion of this work, we learned about
an independent study of random imaginary coupling between the fermions
and bosons by Yuxuan Wang \cite{Wang2019}. Because of the distinction in the fermion-boson coupling
 pairing occurs at higher order
in the expansion in $1/N$. However, our normal state results agree with that of Ref.\cite{Wang2019}.
We are grateful to Y. Wang for sharing his unpublished work with us.

\begin{acknowledgments}
We are grateful to Dimitry Bagrets, Erez Berg, 
Alexander L. Chudnovskiy, J. C. Seamus Davis, Sean A. Hartnoll, Alexey Kamenev, Koenraad
Schalm, Yuxuan Wang, and in particular Andrey V. Chubukov, Steven A. Kivelson and Yoni Schattner for stimulating
discussions. JS was funded by the Gordon and Betty Moore Foundation\textquoteright s
EPiQS Initiative through Grant GBMF4302 while visiting the Geballe
Laboratory for Advanced Materials at Stanford University. IE was supported by NSF grant \# DMR-1608055 at Stanford.
\end{acknowledgments}

\appendix
%dummy comment inserted by tex2lyx to ensure that this paragraph is not empty

\section{Derivation of the Self-Consistency Equations}

After performing the disorder average with the help of the replica
trick, we obtain for the averaged replicated partition function 

\begin{equation}
\overline{Z^{n}}=\int{\cal D}^{n}c^{\dagger}{\cal D}c{\cal D}^{n}\phi e^{-S},
\end{equation}
where the action is of the form
\begin{equation}
S=S_{0}+S_{g}.
\end{equation}
The bare action is given as 
\begin{eqnarray}
S_{0} & = & \sum_{i\sigma a}\int d\tau c_{i\sigma a}^{\dagger}\left(\tau\right)\left(\partial_{\tau}-\mu\right)c_{i\sigma a}\left(\tau\right)+\sum_{ia}\int d\tau\phi_{ia}\left(\tau\right)\left(-\partial_{\tau}^{2}+m_{0}\right)\phi_{ia}\left(\tau\right)\label{eq:S0}
\end{eqnarray}
 while the disorder-average induced interaction term is
\begin{equation}
S_{g}=-\frac{g^{2}}{4N^{2}}\sum_{ijk}\left(\sum_{a\sigma}\int d\tau c_{i\sigma a}^{\dagger}\left(\tau\right)c_{j\sigma a}\left(\tau\right)\phi_{ka}\left(\tau\right)+\sum_{a\sigma}\int d\tau c_{j\sigma a}^{\dagger}\left(\tau\right)c_{i\sigma a}\left(\tau\right)\phi_{ka}\left(\tau\right)\right)^{2},
\end{equation}
a result that can be rewritten as 
\begin{eqnarray}
S_{g} & = & \frac{g^{2}}{2N^{2}}\sum_{ab\sigma\sigma'}\int d\tau d\tau'\sum_{i}^{N}\phi_{ia}\left(\tau\right)\phi_{ib}\left(\tau'\right)\nonumber \\
 & \times & \left[\sum_{i}^{N}c_{i\sigma a}^{\dagger}\left(\tau\right)c_{i\sigma'b}\left(\tau'\right)\sum_{j}^{N}c_{j\sigma'b}^{\dagger}\left(\tau'\right)c_{j\sigma a}\left(\tau\right)\right.\nonumber \\
 & - & \left.\left(\sum_{i}^{N}c_{i\sigma a}^{\dagger}\left(\tau\right)c_{i\sigma'b}^{\dagger}\left(\tau'\right)\right)\left(\sum_{j}^{N}c_{j\sigma'b}\left(\tau'\right)c_{j\sigma a}\left(\tau\right)\right)\right].
\end{eqnarray}
In order to analyze the action we introduce collective variables $G\left(\tau,\tau'\right)$
and Lagrange multiplyer fields $\Sigma\left(\tau,\tau\right)$ 
\begin{eqnarray}
1 & = & \int{\cal D}G\prod_{ab\tau\tau'}\delta\left(NG_{ba,\sigma'\sigma}\left(\tau',\tau\right)-\sum_{i}c_{i\sigma a}^{\dagger}\left(\tau\right)c_{i\sigma'b}\left(\tau'\right)\right)\nonumber \\
 & = & \int{\cal D}G{\cal D}\Sigma e^{\sum_{ab,\sigma\sigma'}\int d\tau d\tau'\left(NG_{ba,\sigma'\sigma}\left(\tau',\tau\right)-\sum_{i}c_{i\sigma a}^{\dagger}\left(\tau\right)c_{i\sigma'b}\left(\tau'\right)\right)\Sigma_{ab,\sigma\sigma'}\left(\tau,\tau'\right)},
\end{eqnarray}
that allow for an efficient decoupling of the interaction terms. Because
of the last term in $S_{g}$ we also include corresponding anomalous
propagators and self energies:
\begin{eqnarray}
1 & = & \int{\cal D}F\prod_{ab\tau\tau'}\delta\left(NF_{ba,\sigma'\sigma}\left(\tau',\tau\right)-\sum_{i}c_{i\sigma a}\left(\tau\right)c_{i\sigma'b}\left(\tau'\right)\right)\nonumber \\
 & = & \int{\cal D}F{\cal D}\Phi^{+}e^{\sum_{ab,\sigma\sigma'}\int d\tau d\tau'\left(NF_{ba,\sigma'\sigma}\left(\tau',\tau\right)-\sum_{i}c_{i\sigma a}\left(\tau\right)c_{i\sigma'b}\left(\tau'\right)\right)\Phi_{ab,\sigma\sigma'}^{+}\left(\tau,\tau'\right)},
\end{eqnarray}
as well as 
\begin{eqnarray}
1 & = & \int{\cal D}F^{+}\prod_{ab\tau\tau'}\delta\left(NF_{ba,\sigma'\sigma}^{+}\left(\tau',\tau\right)-\sum_{i}c_{i\sigma a}^{\dagger}\left(\tau\right)c_{i\sigma'b}^{\dagger}\left(\tau'\right)\right)\nonumber \\
 & = & \int{\cal D}F^{+}{\cal D}\Phi e^{\sum_{ab,\sigma\sigma'}\int d\tau d\tau'\left(NF_{ba,\sigma'\sigma}^{+}\left(\tau',\tau\right)-\sum_{i}c_{i\sigma a}^{\dagger}\left(\tau\right)c_{i\sigma'b}^{\dagger}\left(\tau'\right)\right)\Phi_{ab,\sigma\sigma'}\left(\tau,\tau'\right)}.
\end{eqnarray}
Finally, for the bosonic degrees of freedom we use: 
\begin{eqnarray*}
1 & = & \int{\cal D}D\prod_{ab\tau\tau'}\delta\left(ND_{ab}\left(\tau,\tau'\right)-\sum_{i}\phi_{ia}\left(\tau\right)\phi_{ib}\left(\tau'\right)\right)\\
 & = & \int{\cal D}D{\cal D}\Pi e^{\frac{1}{2}\sum_{ab}\int d\tau d\tau'\left(ND_{ba}\left(\tau',\tau\right)-\sum_{i}\phi_{ia}\left(\tau\right)\phi_{ib}\left(\tau'\right)\right)\Pi_{ab}\left(\tau,\tau'\right)}
\end{eqnarray*}
and obtain an effective action with a sizable amount of integration
variables:
\[
\overline{Z^{n}}=\int{\cal D}G{\cal D}\Sigma{\cal D}F^{+}{\cal D}\Phi^{+}{\cal D}F{\cal D}\Phi{\cal D}D{\cal D}\Pi{\cal D}^{n}c^{\dagger}{\cal D}^{n}c{\cal D}\phi e^{-S}
\]
where the collective action is now given as
\begin{eqnarray}
S & = & \sum_{iab\sigma\sigma'}\int d\tau d\tau'c_{i\sigma a}^{\dagger}\left(\tau\right)\left[\left(\partial_{\tau}-\mu\right)\delta_{ab}\delta_{\sigma\sigma'}\delta\left(\tau-\tau'\right)+\Sigma_{ab,\sigma\sigma'}\left(\tau,\tau'\right)\right]c_{i\sigma'b}\left(\tau'\right)\nonumber \\
 & + & \sum_{iab\sigma\sigma'}\int d\tau d\tau'\left[c_{i\sigma a}^{\dagger}\left(\tau\right)\Phi_{ab,\sigma\sigma'}\left(\tau,\tau'\right)c_{i\sigma'b}^{\dagger}\left(\tau'\right)+c_{i\sigma a}\left(\tau\right)\Phi_{ab,\sigma\sigma'}^{+}\left(\tau,\tau'\right)c_{i\sigma'b}\left(\tau'\right)\right]\\
 & + & \frac{1}{2}\sum_{iab}\int d\tau d\tau'\phi_{ia}\left(\tau\right)\left[\left(-\partial_{\tau}^{2}+m\right)\delta_{ab}\delta\left(\tau-\tau'\right)-\Pi_{ab}\left(\tau,\tau'\right)\right]\phi_{ib}\left(\tau'\right)\nonumber \\
 & - & N\sum_{ab,\sigma\sigma'}\int d\tau d\tau'G_{ba,\sigma'\sigma}\left(\tau',\tau\right)\Sigma_{ab\sigma\sigma'}\left(\tau,\tau'\right)+\frac{N}{2}\sum_{ab}\int d\tau d\tau'D_{ba}\left(\tau',\tau\right)\Pi_{ab}\left(\tau,\tau'\right)\nonumber \\
 & - & N\sum_{ab,\sigma\sigma'}\int d\tau d\tau'F_{ba,\sigma'\sigma}\left(\tau',\tau\right)\Phi_{ab\sigma\sigma'}\left(\tau,\tau'\right)-N\sum_{ab,\sigma\sigma'}\int d\tau d\tau'F_{ba,\sigma'\sigma}^{+}\left(\tau',\tau\right)\Phi_{ab\sigma\sigma'}^{+}\left(\tau,\tau'\right)\nonumber \\
 & + & N\frac{g^{2}}{2}\sum_{ab\sigma\sigma'}\int d\tau d\tau'\left(G_{ab,\sigma\sigma'}\left(\tau,\tau'\right)G_{ba,\sigma'\sigma}\left(\tau',\tau\right)-F_{ab,\sigma\sigma'}^{+}\left(\tau,\tau'\right)F_{ba,\sigma'\sigma}\left(\tau',\tau\right)\right)D_{ab}\left(\tau,\tau'\right).
\end{eqnarray}
We use the Nambu spinor 
\[
\psi_{ia}\left(\tau\right)=\left(c_{i\uparrow a}\left(\tau\right),c_{i\downarrow a}\left(\tau\right),c_{i\uparrow a}^{\dagger}\left(\tau\right),c_{i\downarrow a}^{\dagger}\left(\tau\right)\right)^{T}
\]
and rewrite the first two lines of the previous equation as 
\[
S_{{\rm ferm}}=-\sum_{iab}\int d\tau d\tau'\psi_{ia}^{\dagger}\left(\tau\right)\left(\begin{array}{cc}
G_{0,ab}^{-1}\left(\tau,\tau'\right)-\Sigma_{ab}\left(\tau,\tau'\right) & \Phi_{ab}\left(\tau,\tau'\right)\\
\Phi_{ab}^{+}\left(\tau,\tau'\right) & -\tilde{G}_{0,ba}^{-1}\left(\tau',\tau\right)+\Sigma_{ba}\left(\tau',\tau\right)
\end{array}\right)\psi_{ib}\left(\tau'\right).
\]
 Here we introduced the bare propagators 
\begin{eqnarray*}
G_{0,ab}^{-1}\left(\tau,\tau'\right) & = & -\left(\partial_{\tau}-\mu\right)\delta_{ab}\sigma_{0}\delta\left(\tau-\tau'\right),\\
\tilde{G}_{0,ab}^{-1}\left(\tau,\tau'\right) & = & -\left(\partial_{\tau}+\mu\right)\delta_{ab}\sigma_{0}\delta\left(\tau-\tau'\right).
\end{eqnarray*}
Then we can work with matrices in Nambu space 
\begin{equation}
\hat{G}_{0,ab}^{-1}\left(\tau,\tau'\right)=\left(\begin{array}{cc}
G_{0,ab}^{-1}\left(\tau,\tau'\right) & 0\\
0 & -\tilde{G}_{0,ba}^{-1}\left(\tau',\tau\right)
\end{array}\right)
\end{equation}
and 
\begin{equation}
\hat{\Sigma}_{ab}\left(\tau,\tau'\right)=\left(\begin{array}{cc}
\Sigma_{ab}\left(\tau,\tau'\right) & \Phi_{ab}\left(\tau,\tau'\right)\\
\Phi_{ab}^{+}\left(\tau,\tau'\right) & -\Sigma_{ba}\left(\tau',\tau\right)
\end{array}\right).
\end{equation}
Here $\Sigma_{ab}\left(\tau,\tau'\right)$ and $\Phi_{ab}\left(\tau,\tau'\right)$
etc. are still $2\times2$ matrices in spin space. In addition we
use for the bare phonon propagator
\begin{eqnarray}
D_{0}^{-1}\left(\tau,\tau'\right) & = & \left(-\partial_{\tau}^{2}+m\right)\delta\left(\tau-\tau'\right).
\end{eqnarray}
We can now integrate out the fermions and bosons: 
\begin{eqnarray}
S & = & -N{\rm tr}\log\left(\hat{G}_{0}^{-1}-\hat{\Sigma}\right)+\frac{N}{2}{\rm tr}\log\left(D_{0}^{-1}\left(\tau,\tau'\right)\delta_{ab}-\Pi_{ab}\left(\tau,\tau'\right)\right)\nonumber \\
 & - & N\sum_{ab,\sigma\sigma'}\int d\tau d\tau'G_{ba,\sigma'\sigma}\left(\tau',\tau\right)\Sigma_{ab\sigma\sigma'}\left(\tau,\tau'\right)+\frac{N}{2}\sum_{ab}\int d\tau d\tau'D_{ba}\left(\tau',\tau\right)\Pi_{ab}\left(\tau,\tau'\right)\nonumber \\
 & - & N\sum_{ab,\sigma\sigma'}\int d\tau d\tau'F_{ba,\sigma'\sigma}\left(\tau',\tau\right)\Phi_{ab\sigma\sigma'}\left(\tau,\tau'\right)-N\sum_{ab,\sigma\sigma'}\int d\tau d\tau'F_{ba,\sigma'\sigma}^{+}\left(\tau',\tau\right)\Phi_{ab\sigma\sigma'}^{+}\left(\tau,\tau'\right)\nonumber \\
 & + & N\frac{g^{2}}{2}\sum_{ab\sigma\sigma'}\int d\tau d\tau'\left(G_{ab,\sigma\sigma'}\left(\tau,\tau'\right)G_{ba,\sigma'\sigma}\left(\tau',\tau\right)-F_{ab,\sigma\sigma'}^{+}\left(\tau,\tau'\right)F_{ba,\sigma'\sigma}\left(\tau',\tau\right)\right)D_{ab}\left(\tau,\tau'\right).
\end{eqnarray}
We assume a replica-diagonal structure such that $\overline{Z^{n}}=\overline{Z}^{n}$.
Thus, the average is essentially an annealed one. Now the replica
structure disappears from the action that determines $\overline{Z}$:
\begin{eqnarray}
S & = & -N{\rm tr}\log\left(\hat{G}_{0}^{-1}-\hat{\Sigma}\right)+\frac{N}{2}{\rm tr}\log\left(D_{0}^{-1}-\Pi\right)\nonumber \\
 & - & N\sum_{\sigma\sigma'}\int d\tau d\tau'G_{\sigma'\sigma}\left(\tau',\tau\right)\Sigma_{\sigma\sigma'}\left(\tau,\tau'\right)+\frac{N}{2}\int d\tau d\tau'D\left(\tau',\tau\right)\Pi\left(\tau,\tau'\right)\nonumber \\
 & - & N\sum_{\sigma\sigma'}\int d\tau d\tau'F_{\sigma'\sigma}\left(\tau',\tau\right)\Phi_{\sigma\sigma'}^{+}\left(\tau,\tau'\right)-N\sum_{\sigma\sigma'}\int d\tau d\tau'F_{\sigma'\sigma}^{+}\left(\tau',\tau\right)\Phi_{\sigma\sigma'}\left(\tau,\tau'\right)\nonumber \\
 & + & N\frac{g^{2}}{2}\sum_{\sigma\sigma'}\int d\tau d\tau'\left(G_{\sigma\sigma'}\left(\tau,\tau'\right)G_{\sigma'\sigma}\left(\tau',\tau\right)-F_{\sigma\sigma'}^{+}\left(\tau,\tau'\right)F_{\sigma'\sigma}\left(\tau',\tau\right)\right)D\left(\tau,\tau'\right).
\end{eqnarray}
At large $N$ we can perform the saddle point approximation and obtain
the stationary equations 
\begin{eqnarray}
G\left(\tau,\tau'\right) & = & \left(G_{0}^{-1}-\Sigma\right)_{\tau,\tau'}^{-1},\nonumber \\
D\left(\tau,\tau'\right) & = & \left(D_{0}^{-1}-\Pi\right)_{\tau,\tau'}^{-1},\nonumber \\
\Sigma_{\sigma\sigma'}\left(\tau,\tau'\right) & = & g^{2}G_{\sigma\sigma'}\left(\tau,\tau'\right)D\left(\tau,\tau'\right),\nonumber \\
\Phi_{\sigma\sigma'}\left(\tau,\tau'\right) & = & -g^{2}F_{\sigma\sigma'}\left(\tau',\tau\right)D\left(\tau,\tau'\right),\nonumber \\
\Pi\left(\tau,\tau'\right) & = & -g^{2}\sum_{\sigma\sigma'}\left(G_{\sigma\sigma'}\left(\tau',\tau\right)G_{\sigma'\sigma}\left(\tau,\tau'\right)-F_{\sigma\sigma'}^{+}\left(\tau',\tau\right)F_{\sigma\sigma'}\left(\tau,\tau'\right)\right).
\end{eqnarray}
If we focus on singlet pairing we have $F_{\sigma\sigma'}\left(\tau\right)=F\left(\tau\right)i\sigma_{\sigma\sigma'}^{y}$
and $F_{\sigma\sigma'}^{+}\left(\tau\right)=-F^{+}\left(\tau\right)i\sigma_{\sigma\sigma'}^{y}$. Now
we can rewrite these equations in the usual fashion in $2\times2$
Nambu space with $\left(c_{i\uparrow},c_{i\downarrow}^{\dagger}\right)$
with fermionic Green's function 
\begin{equation}
\hat{G}\left(\omega_{n}\right)^{-1}=i\omega_{n}\tau_{0}+\mu\tau_{3}-\hat{\Sigma}\left(\omega_{n}\right).
\end{equation}
For the bosons we use 
\begin{equation}
D\left(\nu_{n}\right)=\frac{1}{\nu_{n}^{2}+\omega_{0}^{2}+\Pi\left(\nu_{n}\right)}.
\end{equation}
Then, the self energies are given as
\begin{eqnarray}
\hat{\Sigma}\left(\tau\right) & = & g^{2}\tau_{3}\hat{G}\left(\tau\right)\tau_{3}D\left(\tau\right)\nonumber \\
\Pi\left(\tau\right) & = & -g^{2}{\rm tr}\left(\tau_{3}\hat{G}\left(\tau\right)\tau_{3}\hat{G}\left(-\tau\right)\right).
\end{eqnarray}
Those are the coupled equations given above. 

\section{Derivation of the normal-state results}

In this appendix we summarize the derivation of the electron and phonon
propagators for the two normal-state regimes. We start our analysis
with the behavior in the low-temperature quantum critical SYK-regime
and continue with the intermediate temperature impurity-like behavior
at strong coupling. In addition to the analytic derivation we also
present results of the full numerical solution that confirm our analytic
findings in detail.

\subsection{Quantum-critical SYK fixed point: derivation of Eqs.\ref{eq:QC1},
\ref{eq:QC2}, and \ref{eq:omega_r} and numerical results}

We start our analysis at $T=0$ and make the following ansatz for
the fermionic self energy
\begin{equation}
\Sigma\left(\omega\right)=-i\lambda{\rm sign}\left(\omega\right)\left|\omega\right|^{1-2\Delta}.
\end{equation}
To preserve causality, the coefficient $\lambda$ has to be positive.
This is most transparent if one analytically continues this ansatz
to the real frequency axis. Here, causality requires that the retarded
self energy has a negative imaginary part. With ${\rm Im}\Sigma^{R}\left(\epsilon\right)=-\sin\left(\pi\Delta\right)\lambda\left|\epsilon\right|^{\eta}$
follows $\lambda>0$ for $0<\Delta<1$. 

As long as $\Delta>0$ the low-energy fermionic Green's function is
dominated by this singular self energy 
\begin{eqnarray}
G\left(\omega\right) & \approx & -\frac{1}{\Sigma\left(\omega\right)}=-\frac{i}{\lambda}{\rm sign}\left(\omega\right)\left|\omega\right|^{-\left(1-2\Delta\right)}.
\end{eqnarray}
On the real axis this corresponds to the spectral function $A\left(\epsilon\right)=-\frac{1}{\pi}{\rm Im}G^{R}\left(\epsilon\right)=\frac{\sin\left(\pi\Delta\right)\left|\epsilon\right|^{-\left(1-2\Delta\right)}}{\lambda\pi}$
. The bosonic self energy is 
\begin{eqnarray}
\Pi\left(\Omega\right) & = & -2\bar{g}^{2}\int\frac{d\omega}{2\pi}G\left(\omega\right)G\left(\omega+\Omega\right)\nonumber \\
 & = & \frac{2g^{2}}{\lambda^{2}}\int\frac{d\omega}{2\pi}\frac{{\rm sign}\left(\omega\right){\rm sign}\left(\omega+\Omega\right)}{\left|\omega\right|^{1-2\Delta}\left|\omega+\Omega\right|^{1-2\Delta}}
\end{eqnarray}
This bosonic self energy for $\Omega-0$ is ultraviolet divergent if
$\Delta>\frac{1}{4}$, i.e. $\Pi\left(0\right)\propto\Lambda^{4\Delta-1}$
with upper cut-off $\Lambda$. This divergency can be avoided if
we include the full propagator and write 
\begin{eqnarray}
\Pi\left(0\right) & = & -2\bar{g}^{2}\int\frac{d\omega}{2\pi}G\left(\omega\right)^{2}=-2g^{2}\int\frac{d\omega}{2\pi}\left(\frac{1}{i\omega-\Sigma\left(\omega\right)}\right)^{2}\nonumber \\
 & = & \frac{2\Delta-1}{2\Delta^{2}\sin\frac{\pi}{2\Delta}}\bar{g}^{2}\lambda^{-\frac{1}{2\Delta}}.
 \label{eq:AppB Pi(0)}
\end{eqnarray}
Next we analyze the dynamic part $\delta\Pi\left(\Omega\right)=\Pi\left(\Omega\right)-\Pi\left(0\right)$.
It is easiest to do this by first Fourier transforming the propagator
to imaginary time:
\begin{equation}
G\left(\tau\right)=-\frac{\Gamma\left(2\Delta\right)\sin\left(\pi\Delta\right)}{\pi\lambda}\frac{{\rm sign}\left(\tau\right)}{\left|\tau\right|^{2\Delta}}.
\end{equation}
such that the Fourier transform of the phonon self energy is given
as $\Pi\left(\tau\right)=2g^{2}\left(\frac{\Gamma\left(2\Delta\right)\sin\left(\pi\Delta\right)}{\pi\lambda}\right)^{2}\frac{1}{\left|\tau\right|^{4\Delta}}$,
which yields 
\begin{eqnarray*}
\delta\Pi\left(\omega\right) & = & 2\int_{0}^{\infty}\Pi\left(\tau\right)\left(\cos\left(\omega\tau\right)-1\right)d\tau\\
 & = & -\frac{g^{2}}{\lambda^{2}}C_{\Delta}\left|\omega\right|^{4\Delta-1}
\end{eqnarray*}
with coefficient $C_{\Delta}=-8\cos\left(\pi\Delta\right)\sin^{3}\left(\pi\Delta\right)\Gamma\left(2\Delta\right)^{2}\Gamma\left(1-4\Delta\right)/\pi^{2}.$

Now we can analyze the bosonic propagator $D\left(\Omega\right).$ We
can neglect the bare $\Omega^{2}$ term against the singular bosonic
frequency dependence due to the Landau damping. In addition we can
only expect a power law solution if indeed $\omega_{0}^{2}-\Pi\left(0\right)=0$.
If this is the case, it follows for the bosonic propagator
\begin{equation}
D\left(\Omega\right)\approx-\frac{1}{\delta\Pi\left(\Omega\right)}=\frac{\lambda^{2}}{\bar{g}^{2}C_{\Delta}}\left|\Omega\right|^{1-4\Delta}.
\end{equation}
The Fourier transform is $D\left(\tau\right)=\frac{\lambda^{2}}{g^{2}}B_{\Delta}\frac{1}{\left|\tau\right|^{2-4\Delta}}$with
$B_{\Delta}=\frac{\pi\left(1-4\Delta\right)\cos\left(2\pi\Delta\right)}{8\Gamma\left(2\Delta\right)^{2}\cos\left(\pi\Delta\right)\sin^{3}\left(\pi\Delta\right)}$
which gives for the self energy 
\begin{equation}
\Sigma\left(\tau\right)=-\lambda\frac{B_{\Delta}\Gamma\left(2\Delta\right)\sin\left(\pi\Delta\right)}{\pi}\frac{{\rm sign}\left(\tau\right)}{\left|\tau\right|^{2-2\Delta}}.
\end{equation}
Fourier transforming this back to the Matsubara frequency axis finally
yields 
\begin{equation}
\Sigma\left(\omega\right)=-i\lambda A_{\Delta}{\rm sign}\left(\omega\right)\left|\omega\right|^{1-2\Delta}
\end{equation}
with 
\begin{equation}
A_{\Delta}=\frac{4\Delta-1}{2\left(2\Delta-1\right)\left(\sec\left(2\pi\Delta\right)-1\right)}.
\end{equation}
Notice, for the Fourier transforms to be well defined, it must hold
that $\frac{1}{4}<\Delta<\frac{1}{2}$. In order to have a self consistent
solution it must of course hold that $A_{\Delta}=1$. This determines
the exponent $\Delta$ given in Eq. \ref{eq:Delta}. Interestingly,
The coefficient $\lambda$ remains undetermined by this procedure.
However, our solution still relies on the assumption that the renormalized
phonon frequency vanishes at $T=0$. We have not yet determined when
this is the case. We can now always use the freedom and determine
$\lambda$ such that $\omega_{r}\left(T=0\right)=0$, which yields
the condition
\begin{equation}
\lambda=c_{1}g^{4\Delta}\label{eq:lambda}
\end{equation}
in order to generate a critical state for all values of the coupling
constant. The numerical coefficient is 
\begin{equation}
c_{1}=\left(\frac{2\Delta-1}{2\Delta^{2}\sin\frac{\pi}{2\Delta}}\right)^{2\Delta}.\label{eq:c1}
\end{equation}
With $\Delta$ from Eq. \ref{eq:Delta} follows $c_{1}\approx0.8322602114$.
There is one caveat in this argumentation. The relationship between $\Pi(0)$ and $\lambda$ that we used to determine the coefficient $c_1$ relied on the simultaneous knowledge of the low and high-frequency behavior of the fermionic propagator, see Eq.~\ref{eq:AppB Pi(0)}. To address this, we used an expression that  interpolates between the two known limits. Such an approach gives the correct qualitative behavior. Yet the numerical value for $c_{1}$  cannot be reliably determined by such a procedure. To avoid this uncertainty we determined this coefficient from the full numerical solution of the problem that confirms our scaling results in detail; see below. This yields $c_1\approx 1.1547005$ which is somewhat larger than the above estimate. In what follows we will use this result for $c_1$. Notice, all other coefficients of our analysis, such as $C_\Delta$ or $A_\Delta$ can be uniquely determined by the universal low-energy behavior and do not have to be determined numerically.

These results for the phonon frequency allow us to determine the coefficient of the dynamic part of
the boson propagator 
\begin{equation}
\delta\Pi\left(\omega\right)=-c_{3}\left|\frac{\omega}{g^{2}}\right|^{4\Delta-1}
\end{equation}
where $c_{3} =  \frac{C_{\Delta}}{c_{1}^{2}}$.
With $\Delta$ from Eq. \ref{eq:Delta}  and the numerically determined value of $c_1$ follows $c_{3}\approx 0.709618$.

\begin{figure}
\includegraphics[scale=0.6]{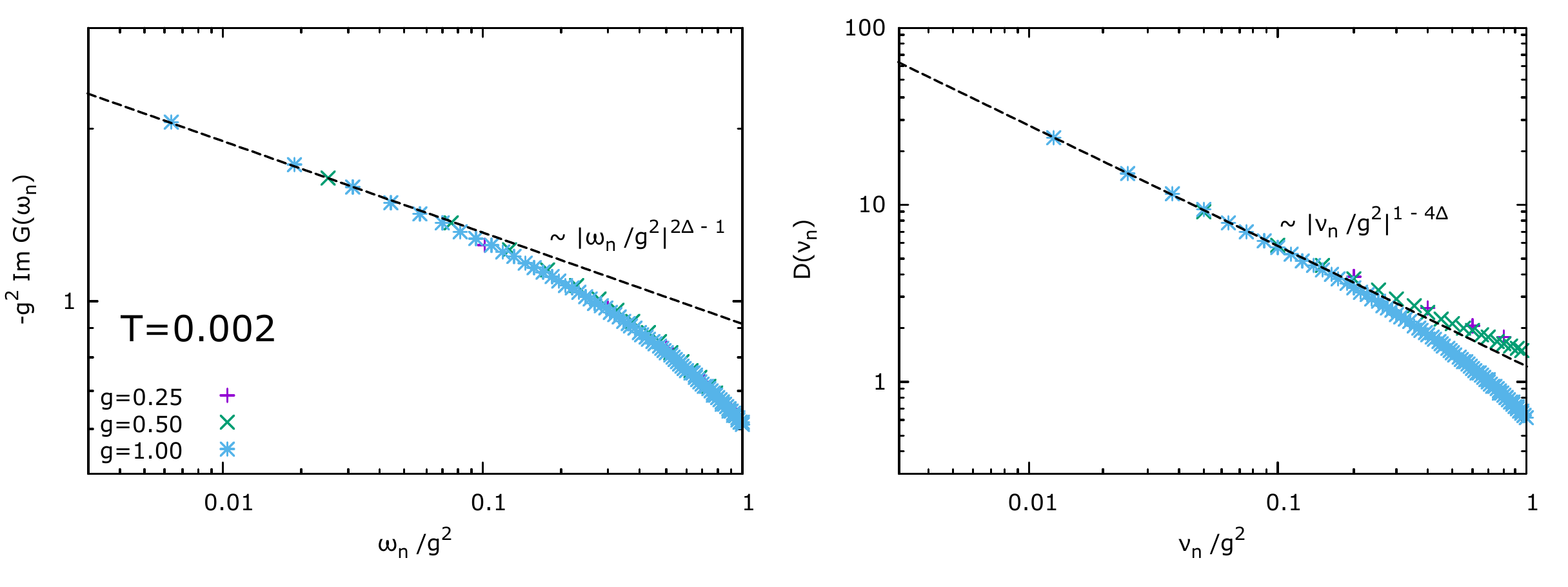}

\caption{Numerical solution of the fermionic (left panel) and bosonic (right
panel) propagators on the imaginary axis in comparison with the analytic
solution given in Eqs.\ref{eq:QC1}, \ref{eq:QC2}. }

\label{numerical SYK_solution}
\end{figure}

This analysis further allows us to determine the temperature dependence
of the phonon frequency, which is determined via 
\begin{equation}
\omega_{r}^{2}\left(T\right)=\omega_{0}^{2}-\Pi\left(T\right),
\end{equation}
where 
\begin{equation}
\Pi\left(T\right)=-2g^{2}T\sum_{n=-\infty}^{\infty}G\left(\omega_{n}\right)^{2}.
\end{equation}
At low but finite temperatures we use for the propagator our result
\begin{equation}
G\left(\omega_{n}\right)=\frac{1}{i\omega_{n}+i\lambda{\rm sign}\left(\omega_{n}\right)\left|\omega_{n}\right|^{1-2\Delta}}.
\end{equation}
Using the Poisson summation formula for fermionic Matsubara sums gives
for the phonon frequency 
\begin{equation}
\omega_{r}^{2}\left(T\right)=\omega_{0}^{2}-2g^{2}\sum_{k=-\infty}^{\infty}\left(-1\right)^{k}\int_{0}^{\infty}\frac{d\omega}{\pi}\frac{\cos\left(\beta\omega k\right)}{\left(\omega+\lambda\omega^{1-2\Delta}\right)^{2}}
\end{equation}
The $k=0$ term corresponds to the $T=0$ result. Thus, it exactly
cancels the bare frequency. The remaining frequency integrals are
ultraviolet convergent even without the bare fermionic propagator
included, which finally gives

\begin{eqnarray}
\omega_{r}^{2}\left(T\right) & = & \frac{4g^{2}}{\lambda^{2}}\sum_{k=1}^{\infty}\left(-1\right)^{k+1}\int_{0}^{\infty}\frac{d\omega}{\pi}\frac{\cos\left(\beta\omega k\right)}{\omega^{2-4\Delta}}\nonumber \\
 & = & c_{2}\left(\frac{T}{g^{2}}\right)^{4\Delta-1},
\end{eqnarray}
with numerical coefficient 
\begin{equation}
c_{2}=\frac{4}{\pi c_{1}^{2}}\sin\left(2\pi\Delta\right)\Gamma\left(4\Delta-1\right)\left(1-2^{2-4\Delta}\right)\zeta\left(4\Delta-1\right),
\end{equation}
where $c_{1}$ was determined numerically, see text below Eq. \ref{eq:c1}. With $\Delta$ from
Eq. \ref{eq:Delta} follows $c_{2}\approx0.561228$.

\begin{figure}
\includegraphics[scale=0.6]{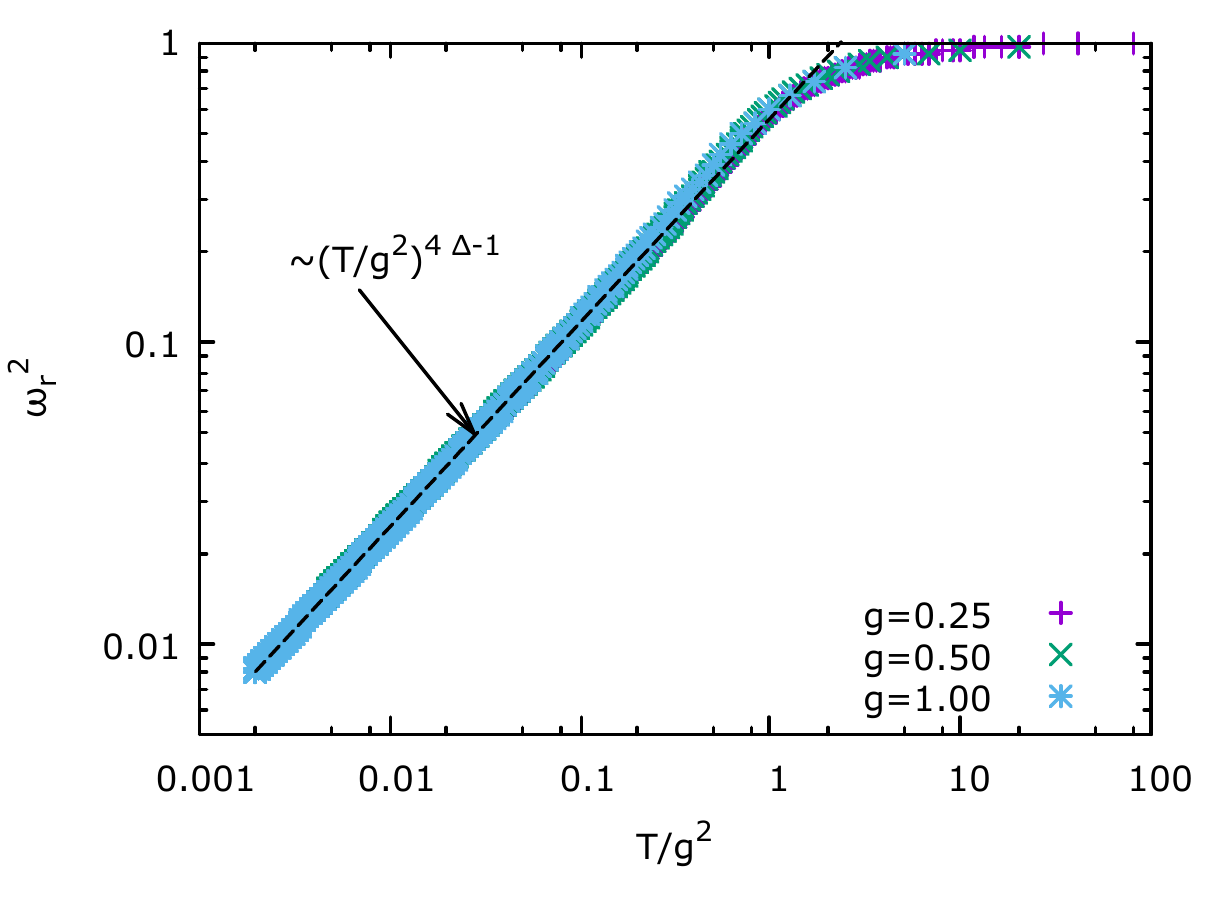}

\caption{Temperature dependence of the renormalized phonon frequency for several
values of the coupling constant $g$ determined from the numerical
solution of the coupled equations and compared with the analytical
expression of Eq.\ref{eq:omega_r}.}

\label{Fig:omegar_SYK}
\end{figure}

We finish this discussion with a comparison of our analytical results
with the numerical solutions of the coupled equations in the normal
state. In Fig.\ref{numerical SYK_solution} we compare the fermionic
and bosonic propagators as function of the imaginary Matsubara frequency
with our analytic solution of Eqs.\ref{eq:QC1}, \ref{eq:QC2}. Finally,
In Fig.\ref{Fig:omegar_SYK} we demonstrate that the phonon frequency
agrees with our analytical result Eq.\ref{eq:omega_r}. In particular
this demonstrates that indeed the phonon frequency is soft for all
values of $g$.

\subsection{Impurity-like fixed point: derivation of Eqs.\ref{eq:SQC1}, \ref{eq:SQC2}
and \ref{eq:Somega_r} and numerical results}

Let us assume that the boson propagator behaves as in Eq.\ref{eq:SQC2}
with renormalized boson frequency $\omega_{r}$, but without additional
dynamic renormalizations due to Landau damping. We further assume
$T\gg\omega_{r}$ something we need to check below to be consistent.
Then follows that the self energy is dominated by the lowest bosonic
Matsubara frequency, i.e. bosons behave as classical impurities:
\begin{eqnarray}
\Sigma\left(\omega_{n}\right) & = & g^{2}T\sum_{n'}D\left(\omega_{n}-\omega_{n'}\right)G\left(\omega_{n'}\right)\nonumber \\
 & = & \frac{g^{2}T}{\omega_{r}^{2}}\frac{1}{i\omega_{n}-\Sigma\left(\omega_{n}\right)}
\end{eqnarray}
This suggests to introduce the energy scale$\Omega_{0}=2\sqrt{\frac{g^{2}T}{\omega_{r}^{2}}}$which
yields 
\begin{equation}
\Sigma\left(\omega_{n}\right)=-i{\rm sign}\left(\omega_{n}\right)\frac{1}{2}\left(\sqrt{\omega_{n}^{2}+\Omega_{0}^{2}}-\left|\omega_{n}\right|\right)
\end{equation}
as solution of the above quadratic equation. For $\left|\omega_{n}\right|\ll\Omega_{0}$
holds $\Sigma\left(\omega_{n}\right)=-i{\rm sign}\left(\omega_{n}\right)\frac{\Omega_{0}}{2}$
while for large frequencies follows $\Sigma\left(\omega_{n}\right)=-i{\rm sign}\left(\omega_{n}\right)\frac{\Omega_{0}^{2}}{4\left|\omega\right|}$.
For the fermionic Green's function follows then Eq.\ref{eq:SQC1}.
Next we determine the bosonic self energy for this problem: 
\begin{equation}
\Pi\left(\omega_{n}\right)=-2g^{2}T\sum_{n'}G\left(\omega_{n'}\right)G\left(\omega_{n'}+\omega_{n}\right).
\end{equation}
Let us first determine the zero frequency part
\begin{eqnarray}
\Pi\left(0\right) & = & -2g^{2}T\sum_{n'}G\left(\omega_{n'}\right)^{2}\nonumber \\
 & = & 8g^{2}T\sum_{n'}\frac{1}{\left(\sqrt{\omega_{n}^{2}+\Omega_{0}^{2}}+\left|\omega_{n}\right|\right)^{2}}
\end{eqnarray}
Let us try to determine $\Omega_{0}$ from the condition that the
boson frequency goes to zero as $T$ is extrapolated to $T=0$. Formally
we can just require that $\Pi\left(0\right)=\omega_{0}^{2}$ at $T=0.$
Then we have `
\begin{eqnarray}
\Pi\left(0\right) & = & 8g^{2}\int_{0}^{\infty}\frac{d\omega}{\pi}\frac{1}{\left(\sqrt{\omega^{2}+\Omega_{0}^{2}}+\omega\right)^{2}}\nonumber \\
 & = & \frac{16g^{2}}{3\pi\Omega_{0}}
\end{eqnarray}
This yields $\Omega_{0}=\frac{16}{3\pi}g^{2}$. Combining both expressions that
we obtained for $\Omega_{0}$ can be used to determine the phonon frequency and
gives rise to our result Eq.\ref{eq:Somega_r}. The assumption of
classical bosons was $T\gg\omega_{r}$ which implies $T\gg g^{-2}$,
consistent in the strong coupling limit. In addition, as long as $T\ll g^{2}$
we also have $T\ll\Omega_{0}$ and the evaluation of the above fermionic
Matsubara sum in the zero-temperature limit is justified. The frequency
dependence of the self energy for $\omega\ll g^{2}$ is then$\Sigma\left(\omega_{n}\right)=-i{\rm sign}\left(\omega_{n}\right)\frac{8}{3\pi}g^{2}$. 

\begin{figure}
\includegraphics[scale=0.5]{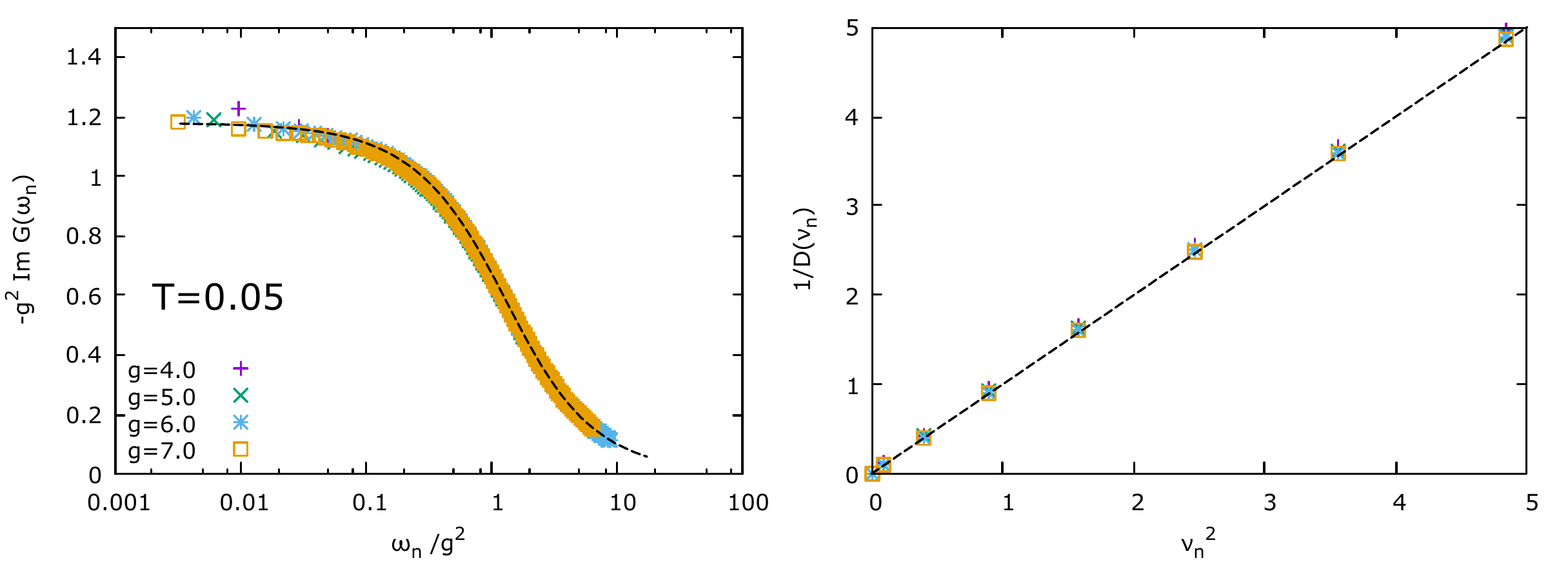}

\caption{Numerical solution of the fermionic (left panel) and bosonic (right
panel) propagators on the imaginary axis in comparison with the analytic
solution given in Eqs.\ref{eq:SQC1}, \ref{eq:SQC2}. }

\label{numerical imp_solution}
\end{figure}

For consistency we have to check that we can indeed ignore the frequency
dependence of the bosonic self energy. The only scale that enters
the fermionic propagator is $\Omega_{0}$. In the relevant limit $T\ll\Omega_{0}$
the fermions are essentially at zero temperature, where
\begin{eqnarray*}
\delta\Pi\left(\omega\right) & = & 2\int_{0}^{\infty}d\tau\Pi\left(\tau\right)\left(\cos\left(\omega\tau\right)-1\right)\\
 & = & -4g^{2}\int_{0}^{\infty}d\tau G\left(\tau\right)G\left(-\tau\right)\left(\cos\left(\omega\tau\right)-1\right)
\end{eqnarray*}
 The Fourier transform of the fermionic propagator can be determined
analytically and expressed in terms of modified Bessel functions and
the modified Struve function. For our purposes it suffices to analyze
the short and long time limit:
\begin{equation}
G\left(\tau\right)={\rm sign}\left(\tau\right)\times\left\{ \begin{array}{cc}
\frac{1}{\Omega_{0}\left|\tau\right|}\,\, & {\rm if}\,\,\left|\tau\right|\gg\Omega_{0}^{-1}\\
\frac{1}{2}-\frac{2}{3\pi}\left|\tau\right|\Omega_{0}\,\, & {\rm if}\,\,\left|\tau\right|\ll\Omega_{0}^{-1}
\end{array}\right.,
\end{equation}
which yields
\begin{eqnarray*}
\delta\Pi\left(\omega\right) & \approx- & \frac{\left|\omega\right|}{\Omega_{0}}.
\end{eqnarray*}
This Landau damping term is negligible compared to $\omega_{n}^{2}$
for $T\gg g^{-2}$. Thus, we can indeed approximate the bosonic propagator
by Eq.\ref{eq:SQC2}.

\begin{figure}
\includegraphics[scale=0.6]{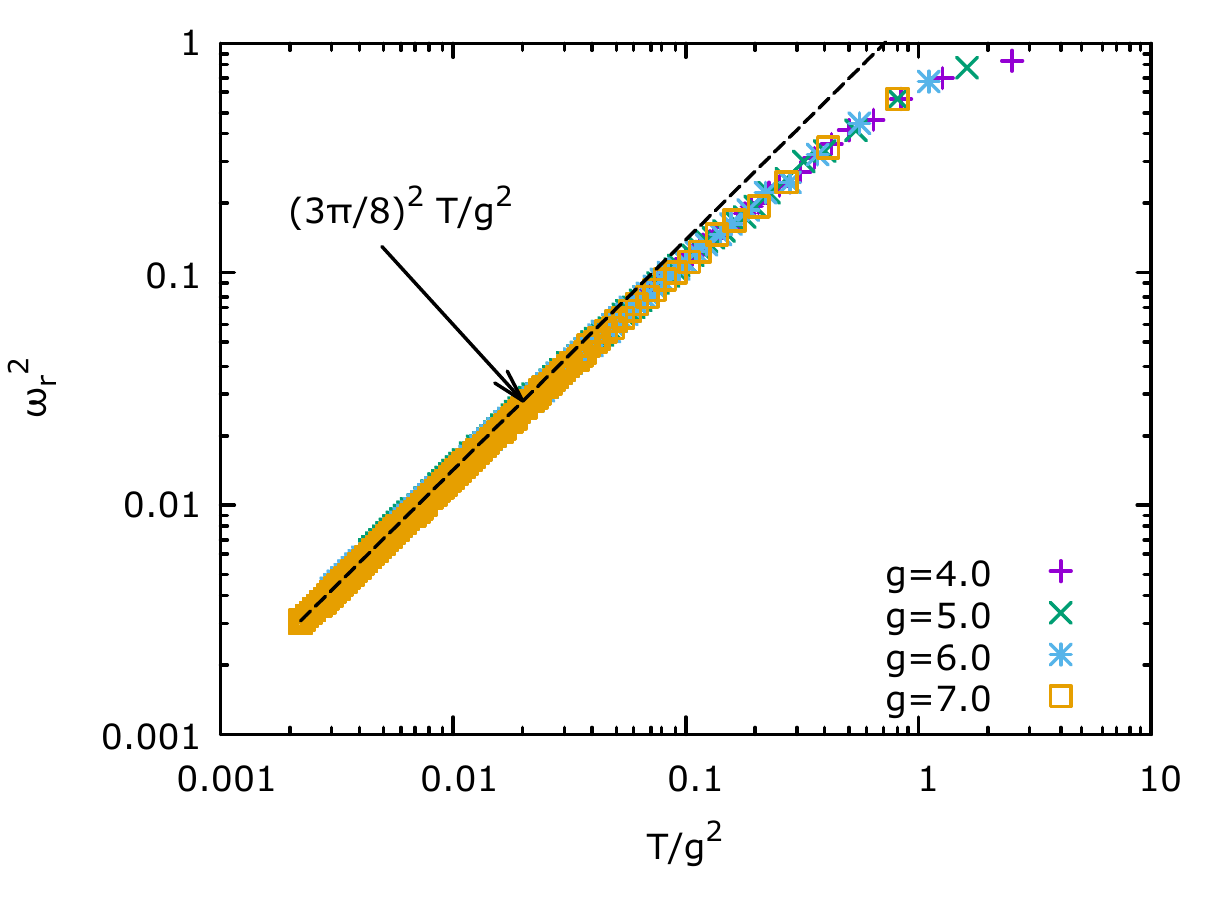}

\caption{Temperature dependence of the renormalized phonon frequency for several
values of the coupling constant $g$ determined from the numerical
solution of the coupled equations and compared with the analytical
expression of Eq.\ref{eq:Somega_r}.}

\label{Fig:omegar_imp}
\end{figure}

We finish this discussion with a comparison of our analytical results
with the numerical solutions of the coupled equations in the normal
state. In Fig.\ref{numerical imp_solution} we compare the fermionic
and bosonic propagators as function of the imaginary Matsubara frequency
with our analytic solution of Eqs.\ref{eq:SQC1}, \ref{eq:SQC2}.
Finally, In Fig.\ref{Fig:omegar_imp} we demonstrate that the phonon
frequency agrees with our analytical result Eq.\ref{eq:Somega_r}. 

\section{On the role of distinct fermion and boson modes}

The ratio $m=M/N$ changes the relative importance of the fermion
and boson self energies. Changing the ratio $m$ of the number of boson
and fermion flavors does not affect the overall behavior of Eqs.\ref{eq:GOE1}
and \ref{eq:QC2}. The exponent $\Delta$ changes continuously from
$\Delta\left(m\rightarrow0\right)\rightarrow1/2$ to $\Delta\left(m\rightarrow\infty\right)\rightarrow1/4$.
The phonon softening follows formally still Eq.\ref{eq:omega_r},
yet the temperature scale below which this powerlaw softening occurs
depends sensitively on the relative importance of the phonon and electron
renormalizations. If phonon self energy effects dominate ($m\ll1$)
we find $\omega_{r}^{2}=\frac{m}{4}\pi^{2}\log2\left(T/g^{2}\right)^{1-\frac{m}{2}}$,
i.e. phonons are soft below a very large temperature $T^{*}\sim g^{2}/m^{1-\frac{m}{2}}$.
In the opposite limit, of large $m$, i.e. relatively negligible phonon
self energy, holds that $\omega_{r}^{2}\approx\left(\frac{T}{g^{2}}\right)^{\sqrt{\frac{2}{\pi m}}}$and
the temperature window below phonon softening takes place is exponentially
small $T^{*}\sim g^{2}e^{-\sqrt{\frac{\pi m}{2}}}$. 
\end{document}